\documentclass[11pt]{amsart}

\usepackage{amsmath}
\usepackage{amssymb}
\usepackage{epsfig}
\usepackage{comment}
\usepackage{subcaption}
\usepackage[section]{placeins}
\usepackage{mathrsfs}
\usepackage{graphicx}
\usepackage[notrig]{physics}
\usepackage{units}
\usepackage{xcolor}
\usepackage{algorithm}
\usepackage{algorithmic}

\usepackage{blindtext}
\usepackage{geometry}

\newtheorem{thm}{Theorem}[section]

\theoremstyle{definition}

\newtheorem{example}[thm]{Example}

\theoremstyle{remark}

\begin{document}

\title[Model-free Data-Driven inference]
{
    Model-free Data-Driven inference \\ in computational mechanics
}
\author{E.~Prume${}^1$, S.~Reese${}^1$ and M.~Ortiz${}^{2,3}$}

\address
{
    ${}^1$ Institute of Applied Mechanics, RWTH Aachen University,
    Mies-van-der-Rohe-Str.~1, D-52074 Aachen, Germany. \\ \newline\indent
    ${}^2$Hausdorff Center for Mathematics, Universit\"at Bonn, Endenicher Allee 60, 53115 Bonn, Germany. \\ \newline\indent
    ${}^3$Division of Engineering and Applied Science, California Institute of Technology, 1200 E.~California Blvd., Pasadena, CA 91125, USA.
}

\begin{abstract}
We extend the model-free Data-Driven computing paradigm to solids and structures that are stochastic due to intrinsic randomness in the material behavior. The behavior of such materials is characterized by a likelihood measure instead of a constitutive relation. We specifically assume that the material likelihood measure is known only through an empirical point-data set in material or phase space. The state of the solid or structure is additionally subject to compatibility and equilibrium constraints. The problem is then to infer the likelihood of a given structural outcome of interest. In this work, we present a Data-Driven method of inference that determines likelihoods of outcomes from the empirical material data and that requires no material or prior modeling. In particular, the computation of expectations is reduced to explicit sums over local material data sets and to quadratures over admissible states, i.~e., states satisfying compatibility and equilibrium. The complexity of the material data-set sums is linear in the number of data points and in the number of members in the structure. Efficient population annealing procedures and fast search algorithms for accelerating the calculations are presented. The scope, cost and convergence properties of the method are assessed with the aid selected applications and benchmark tests.
\end{abstract}

\maketitle

\section{Introduction}

Some classes of materials exhibit behavior that is intrinsically random. A compelling example is set forth by brittle materials with random tensile strength \cite{Bower:2010}. Structural members made of such materials may fail -- and lose their load-bearing capacity altogether -- at tensile stresses that are sensitive to critical flaws in the material. Since such flaws have random geometry, the tensile strength of the material is likewise random (cf., e.~g., refs.~\cite{Reynolds:1974, Bennett:1983} concerning the example of carbon fibers). In addition, empirical material data may be noisy due to experimental measurement error and scatter. This observational noise may be reduced by improving the precision of the experimental measurements. By contrast, the intrinsic randomness of the material is not eliminated by removing observational noise.

\begin{figure}[H]
	\begin{subfigure}{0.45\textwidth}\caption{} \includegraphics[width=0.99\linewidth]{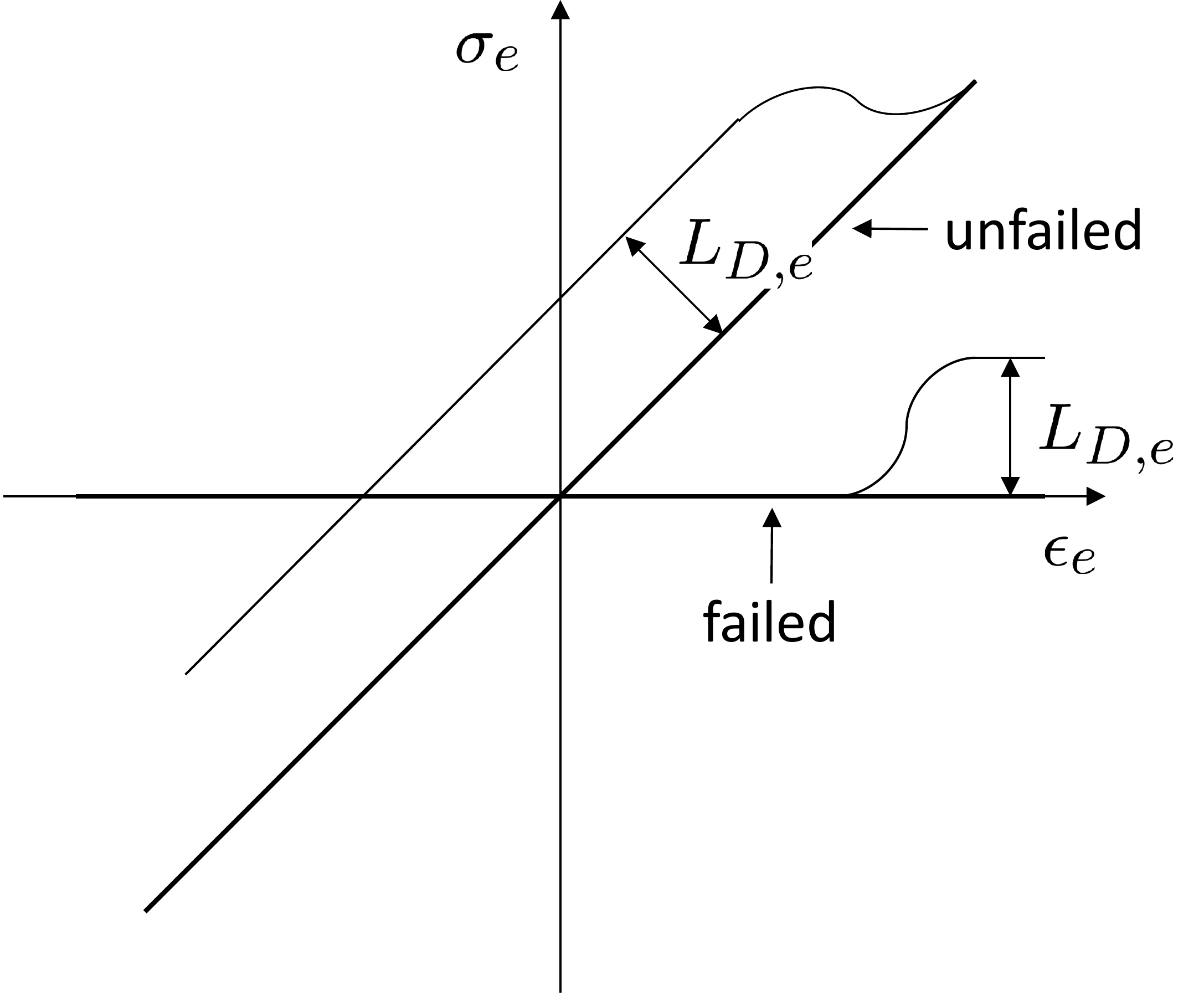}
	\end{subfigure}
    $\quad$
	\begin{subfigure}{0.45\textwidth}\caption{} \includegraphics[width=0.99\linewidth]{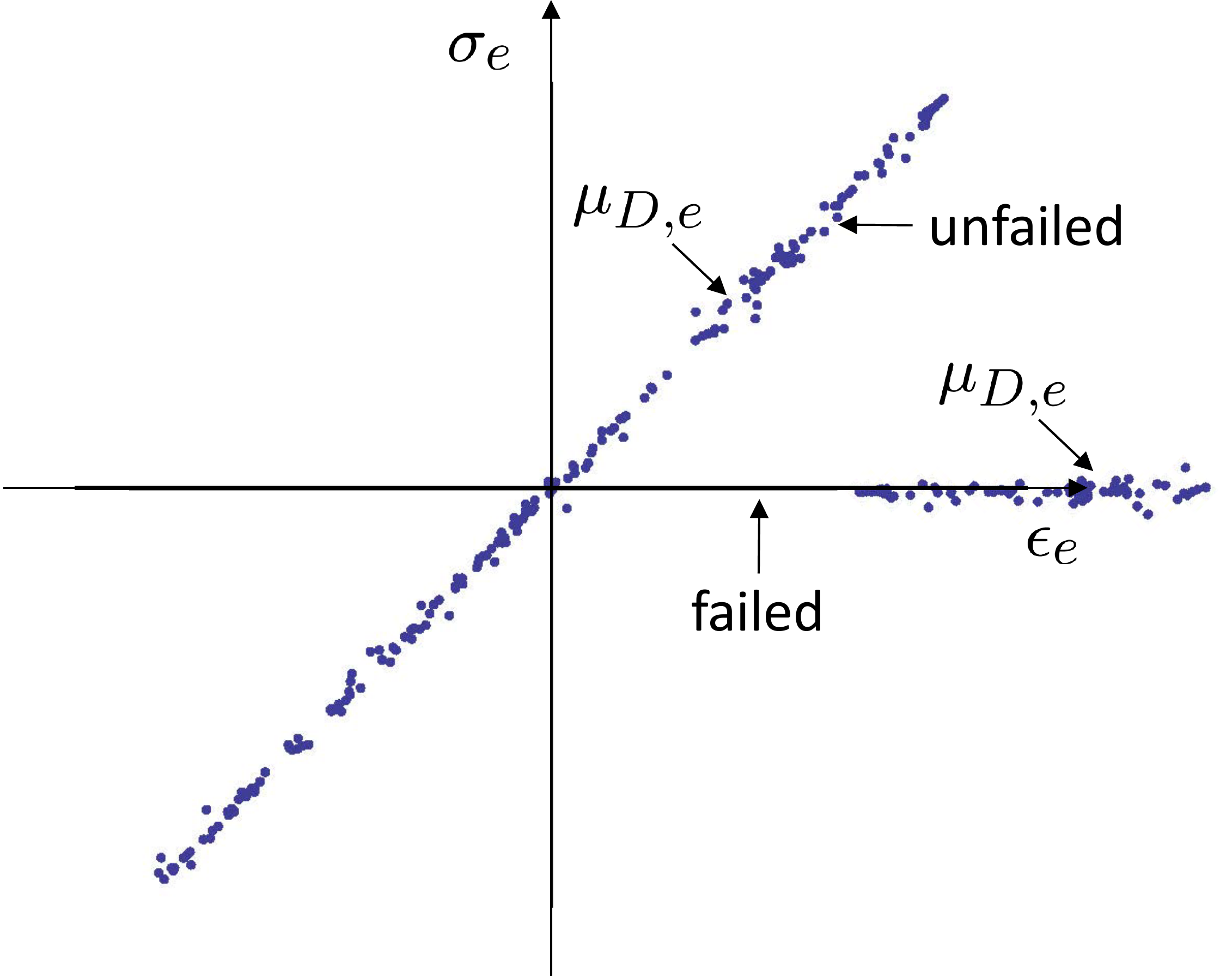}
	\end{subfigure}
    \caption{Brittle material with random tensile strength, member $e$ of the structure.
    a) Local material likelihood function $L_{D,e}$ providing the likelihood that the local material state $(\epsilon_e,\sigma_e)$ be on the unfailed and failed branches of the local material set, respectively. b) Empirical local likelihood measure $\mu_{D,e}$ sampled from $L_{D,e}$.} \label{uCXq6P}
\end{figure}

The behavior of random materials is characterized by a likelihood measure instead of a constitutive relation. In principle, any local material state $(\epsilon_e,\sigma_e)$ is possible, albeit with varying likelihood. For instance, for a brittle material with random tensile strength undergoing monotonically increasing strains, the stress may take values in anyone of two branches, corresponding to a failed or un-failed material, with respective likelihoods determined by the distribution of tensile strength, Fig.~\ref{uCXq6P}a.

In addition, local stresses and strains in solids and structures are subject to compatibility and equilibrium constraints. These field constraints are known exactly, though the loading may also be random in some applications. The inference problem is then to determine the likelihood of structural outcomes of interest when the states of the structure are required to be admissible, in the sense of satisfying compatibility and equilibrium, and the material is characterized by a given material likelihood measure.

However, the requisite material likelihood measures are often unknown exactly and characterized only partially through empirical material data, Fig.~\ref{uCXq6P}b. A common response to that challenge is to model the empirical material data, e.~g., by assuming the existence of an underlying reduced manifold in phase space masked by noise obeying a prespecified prior distribution. However, in general no such reduced manifold may exist and the form of the prior may be unknown and subject to surmise, which inevitably introduces modeling bias and uncertainty and may prevent convergence with respect to the empirical data altogether \cite{Owhadi:2015}.

For instance, for brittle materials with random tensile strength the material states may occupy one of two branches, corresponding to a failed or un-failed material, and no single-valued stress-strain relation exists, Fig.~\ref{uCXq6P}a. In addition, in approaches based on Bayesian statistics (cf., e.~g., \cite{Dashti:2017} for a review) it is common to presume a certain class of priors, most commonly Gaussian \cite{Knapik:2011, Knapik:2016}, which may not contain the actual likelihood function of the material. For instance, if the tensile strength of a brittle material obeys a Weibull distribution (cf., e.~g., \cite{Bader:1993, Naito:2008} for the example of carbon fibers), methods assuming a Gaussian prior inevitably converge to the wrong answer with increasing sample size.

In this work, we present an alternative method of inference that determines likelihoods of outcomes directly from the empirical material data and requires no material or prior-distribution {\sl ans\"atze}. We begin by 'thermalizing' the data, which effectively replaces the empirical material likelihood by a sum of Gaussians of a certain width \cite{Conti:2021}. The thermalized likelihoods also follow variationally by minimizing a regularized Kullback-Leibler divergence \cite{Kullback:1951, Kullback:1959, Pinski:2015} from the empirical material likelihood. Likewise, an optimal annealing schedule follows variationally by minimizing the regularized Kullback-Leibler divergence with respect to temperature. The sequence of thermalized likelihoods thus obtained is then used to approximate expectations of outcomes. In this manner, the computation of expectations is reduced to explicit sums over local material data sets and quadratures over admissible states, i.~e., states satisfying compatibility and equilibrium. When restricted to the computation of maximum-likelihood outcomes, the present approach reduces to the minimum-distance (min-dist) Data-Driven method of \cite{kirchdoerfer2016data} in the deterministic case with uniform data convergence, and to the maximum-entropy (max-ent) Data-Driven method of \cite{kirchdoerfer2017data} in the case of noisy data with outliers.

It bears emphasis that, in the present Data-Driven approach, expectations are computed directly from the material data and that no material model or prior distribution need to be contrived at any point of the calculations. The complexity of the Data-Driven inference calculations is linear in the number of material data points and in the number of members in the structure or Gauss points in the solid. We formulate efficient population annealing procedures and fast-search algorithms for accelerating calculations further. Specifically, the population annealing algorithm reduces the complexity of the calculations and it renders it linear in the size of the material data-point population. The calculations are further accelerated by recourse to a hierarchical $k$-means algorithm for the computation of the annealing energy with limits on the number backtracking operations.

The convergence properties of the method with respect to the empirical data are assessed with the aid of selected applications and benchmark tests, including a standard verification test based on Gaussian material data, brittle materials exhibiting random tensile strength, and a simple lightweight space structure. Robust convergence of posterior distributions is obtained in these tests, consistent with quantitative error estimates derived from analysis \cite{Conti:2021, Conti:2022}. The ability of the proposed Data-Driven method of inference to deal effectively with general material data sets and complex behavior without need for models, hypotheses or assumptions is quite remarkable. The numerical tests additionally attest to feasibility of the calculations and the effectiveness of acceleration methods such as population annealing and hierarchical $k$-means searches.

\section{The inference problem}
\label{M2Xt7P}

We consider finite-dimensional systems comprising $m$ components whose state is characterized by two work-conjugate fields $\epsilon \equiv \{\epsilon_e \in \mathbb{R}^d,\ e=1,\dots,m\}$ and $\sigma \equiv \{\sigma_e \in \mathbb{R}^d,\ e=1,\dots,m\}$. We refer to the space of pairs $Z_e = \{z_e \equiv (\epsilon_e, \sigma_e) \in \mathbb{R}^d \times \mathbb{R}^d\}$ as the {\sl local phase space} of component $e$, and $Z = Z_1 \times \cdots \times Z_m = \mathbb{R}^{N\times N}$, $N = m d$, as the {\sl global phase space} of the system. We suppose that a suitable norm is defined in $Z$, e.~g.,
\begin{equation}\label{Poyet2}
    \| z \|
    =
    \| (\epsilon,\sigma) \|
    =
    \left(
    \sum_{e=1}^m
        w_e
        \Big(
            {\mathbb{C}}_e \epsilon_e \cdot \epsilon_e
            +
            {\mathbb{C}}_e^{-1} \sigma_e \cdot \sigma_e
        \Big)
    \right)^{1/2} ,
\end{equation}
where $w_e > 0$ are weights and $\mathbb C_e\in \mathbb R^{d\times d}_{{\rm sym},+}$ are positive definite symmetric matrices, $e=1,\dots,m$.

\begin{example}[Trusses]\label{nBpq7d}
We illustrate the essential structure of discrete field theories by means of the simple example of truss structures. Trusses are assemblies of bars that deform in uniaxial tension or compression. The bars are articulated at common joints, or nodes, that act as hinges, i.~e., cannot transmit moments. Trusses are examples of connected networks that obey conservation laws. Other examples in the same class include electrical circuits, pipeline networks, traffic networks, and others.

The material behavior of a bar $e$ is characterized by a particularly simple relation between uniaxial strain $\epsilon_e$ and uniaxial stress $\sigma_e$. Thus, in this case the local phase spaces are $Z_e = \mathbb{R} \times \mathbb{R}$. These local states are subject to the following laws:

i) Compatibility: Suppose that bar $e$ is connected to nodes $a$ and $b$. Then, the strain in the bar is
\begin{equation}
    \epsilon_e = \frac{u_b - u_a}{L_e} \cdot d_e ,
\end{equation}
where $L_e$ is the length of the bar and $d_e$ is the unit vector pointing from $a$ to $b$.

ii) Equilibrium: Let $S_a$ be the star of an unconstrained node $a$, i.~e., the collection of members connected to $a$. Then, we must have
\begin{equation}
    \sum_{e\in S_a} \sigma_e d_e A_e + f_a = 0,
\end{equation}
where $A_e$ is the cross-sectional area of bar $e$ and $f_a$ is the force applied to node $a$. \hfill$\square$
\end{example}

\subsection{The constraint set of admissible states}\label{sec:Cset}

The above example shows that the state of the system is subject to linear constraints of the general form
\begin{subequations}\label{9qHWzU}
\begin{align}
    &
    \sum_{e=1}^m w_e B_e^T \sigma_e = f ,
    \\ &
    \epsilon_e = B_e u + g_e , \quad e = 1,\dots m ,
\end{align}
\end{subequations}
where $u \in \mathbb{R}^{n}$ is the array of degrees of freedom of the system, $w_e$ are positive weights, $B_e \in \mathbb{R}^{d \times n}$ is a discrete gradient operator, $B_e^T$ is a discrete divergence operator, $f \in \mathbb{R}^n$ is a force array resulting from distributed sources and Neumann boundary conditions and the arrays $g_e \in \mathbb{R}^{d}$ follow from Dirichlet boundary conditions. The constraints (\ref{9qHWzU}) are material independent and define an affine subspace $E$ of $Z$, the {\sl constraint set}. The constraint set $E$ encodes all the data of the problem, including geometry, loading and boundary conditions. The constraints (\ref{9qHWzU}) can also be expressed in matrix form as
\begin{subequations}\label{vZftZT}
\begin{align}
    &
    B^T \tau = f
    \\ &
    \epsilon = B u + g ,
\end{align}
\end{subequations}
with $B = (B_1, \dots, B_m) \in \mathbb{R}^{N \times n}$, $\epsilon = (\epsilon_1, \dots, \epsilon_m)  \in \mathbb{R}^N$, $\tau = (w_1 \sigma_1,\dots,$ $w_m \sigma_m)$  $\in$ $\mathbb{R}^N$ and $g = (g_1, \dots, g_m)  \in \mathbb{R}^N$.

We note that the affine space $E$ defined by the constraints (\ref{vZftZT}) is a translate of the linear space $E_0$ defined by the homogeneous constraints
\begin{subequations}
\begin{align}
    &   \label{qmS3Q5}
    B^T \tau = 0
    \\ &  \label{jJ9B6T}
    \epsilon = B u .
\end{align}
\end{subequations}
Evidently, $E_0 = E_\epsilon \times E_\sigma$, where $E_\epsilon$ is the linear space defined by (\ref{jJ9B6T}) and $E_\sigma$ is the linear space defined by (\ref{qmS3Q5}). Therefore, we have
\begin{equation}
    {\rm dim}(E_0)
    =
    {\rm dim}(E_\epsilon)
    +
    {\rm dim}(E_\sigma)
    =
    {\rm dim}({\rm Im}(B)) + {\rm dim}({\rm Ker}(B^T))
    =
    N .
\end{equation}
Since $Z = \mathbb{R}^N \times \mathbb{R}^N$, it follows that the constraint set $E$ is an affine subspace of $Z$ of dimension $N$ and co-dimension $N$.

This observation  characterizes the structure and dimensionality of the phase space $Z$ and of the subspace $E$ of all admissible states in $Z$, or constraint set, i.~e., the set of all the states that satisfy the conservation laws (\ref{vZftZT}). To wit, the phase space has the conventional dual structure $Z = \mathbb{R}^N \times \mathbb{R}^N$ of work-conjugate pairs, whereas the constraint set $E$ is a linear subspace of $Z$ of dimension $N$ and co-dimension $N$. This structure is particular to systems obeying conservation laws and sets them apart from other applications in data science where the data is heterogeneous and unstructured.

\subsection{Material characterization}

We assume that the material behavior is inherently random and characterized by a positive, continuous, likelihood function $L_D(y)$ representing the likelihood of observing a material state $y\in Z$, cf.~Fig.~\ref{uCXq6P}a. Since material behavior is not localized to a bounded region of phase space in general, the material likelihood function $L_D$ contains infinite mass and cannot be normalized to define a probability density. The function $\Phi_D(y) = - \log L_D(y)$ is the corresponding {\sl potential}.

\begin{example}[Local material behavior] \label{dzeXXX}{\rm
Material behavior is often {\sl local} and can be characterized over each member phase space $Z_e = \mathbb{R}^d \times \mathbb{R}^d$ by a local material likelihood function $L_{D,e} \in \mathcal{M}(Z_e)$. Assuming that the behavior of the members is independent, then the {\sl global} likelihood function is
\begin{equation}
    L_D = L_{D,1} \times \cdots \times L_{D,m} ,
\end{equation}
and the {\sl global} potential is
\begin{equation}
    \Phi_D = \Phi_{D,1} + \cdots + \Phi_{D,m} .
\end{equation}
Likelihood functions need not be integrable and do not define a probability density in general. For instance, suppose that the local behavior is characterized by a sliding Gaussian of the form, cf.~Fig.~\ref{uCXq6P}a.,
\begin{equation}
    L_{D,e}(y_e)
    =
    \exp\Big( - \frac{w_e}{2s^2} \| \sigma_e - \mathbb{C}_e \epsilon_e \|^2\Big)
\end{equation}
where $y_e = (\epsilon_e, \sigma_e) \in Z_e$ and the parameter $s>0$ measures the width of the distribution. The global likelihood function is, then,
\begin{equation}\label{nJFGU9}
    L_{D}(y)
    =
    \exp
    \Big(
        -
        \frac{1}{2s^2}
        \sum_{e=1}^m w_e \| \sigma_e - \mathbb{C}_e \epsilon_e \|^2
    \Big) ,
    \quad
    \| \sigma_e \|^2 = \mathbb{C}_e^{-1} \sigma_e \cdot \sigma_e ,
\end{equation}
$y = (\epsilon,\sigma)$. Equivalently,
\begin{equation}\label{0VUrvV}
    L_{D}(y)
    =
    \exp
    \Big(
        -
        \frac{1}{2s^2}
        \Big(
            \| \epsilon \|^2 + \| \sigma \|^2 - 2 \sigma \cdot \epsilon
        \Big)
    \Big) ,
\end{equation}
with
\begin{equation}
    \| \epsilon \|^2
    =
    \sum_{e=1}^m
         w_e \mathbb{C}_e \epsilon_e \cdot \epsilon_e ,
    \quad
    \| \sigma \|^2
    =
    \sum_{e=1}^m
         w_e \mathbb{C}_e^{-1} \sigma_e \cdot \sigma_e ,
    \quad
    \sigma \cdot \epsilon = \sum_{e=1}^m w_e \sigma_e \cdot \epsilon_e ,
\end{equation}
or
\begin{equation}
    L_{D}(y)
    =
    \exp
    \Big(
        -
        \frac{1}{2s^2}
        \Big(
            \| y \|^2 - 2 \sigma \cdot \epsilon
        \Big)
    \Big)
\end{equation}
in the norm (\ref{Poyet2}). Evidently, $L_{D,e}$ decays away from the centerline $D_e = \{ \sigma_e - \mathbb{C}_e \epsilon_e \}$ but is constant along the line. Hence,
\begin{equation}
    \int_{Z_e} L_{D,e}(y_e) \, dy_e = +\infty ,
\end{equation}
and neither $L_{D,e}(y_e)$ nor $L_D(y)$ can be normalized.
\hfill$\square$}
\end{example}

\subsection{Classical inference}

If the constraint set $E$ and the material likelihood function $L_D$ are fully known, , cf.~Fig.~\ref{uCXq6P}a, the expectation of any quantity of interest, represented by a bounded continuous function $f \in C_b(Z)$, is simply given by


\begin{equation}\label{K2mXSg}
    \mathbb{E}[f]
    =
    \frac{
        \int_E
            f(z) L_D(z)
        \, dz 
    }
    {
        \int_E
            L_D(z)
        \, dz 
    } ,
\end{equation}


which fully characterizes the response of the system in the sense of probability.

\begin{example}[Random trusses]\label{suO4Ci}
Consider a truss such as defined in Example~\ref{nBpq7d}, with material behavior characterized by the likelihood function (\ref{nJFGU9}). In addition, we metrize phase space by means of the Euclidean norm (\ref{Poyet2}). We begin by parameterizing the constraint space $E$. Recall that $m$ is the number of members of the truss, $Z = \mathbb{R}^m \times \mathbb{R}^m$ is the phase space and $n < m$ the number of unconstrained degrees of freedom. Let $l = m-n$ and $A \in \mathbb{R}^{m\times l}$ the matrix whose columns define a basis of ${\rm Ker}(B^T)$. Further, we define the matrix of Gauss point volumes as $\mathbb{W}={\rm diag}(w_1,\dots,w_m)$. Suppose non-degeneracy in the sense that $\sigma$ satisfies the equilibrium condition $B^T\mathbb{W}\sigma = 0$ if and only if there is $v \in \mathbb{R}^l$ such that
\begin{equation}\label{RYaYcU}
    \sigma = \mathbb{W}^{-1}A v .
\end{equation}
We may thus regard $v$ as a discrete Airy potential and $A$ as a discrete Airy operator. In addition, suppose that $\epsilon$ satisfies the compatibility condition $A^T \epsilon = 0$ if and only if there are displacements $u$ such that $\epsilon = B u$. Then, the constraint set $E_0$ through the origin, corresponding to $f=0$ and $g=0$, admits the representation
\begin{equation}
    E_0
    =
    \{
    (\epsilon,\sigma) \in Z = \mathbb{R}^m \times \mathbb{R}^m
    \, : \,
    \epsilon = B u,\ u \in \mathbb{R}^n;
    \ \sigma = \mathbb{W}^{-1}A v,\ v \in \mathbb{R}^l \} .
\end{equation}
The general constraint set is then the translation
\begin{equation}
    E = z_0 + E_0,
    \quad
    z_0 = (\epsilon_0,\sigma_0) ,
    \quad
    \epsilon_0 = g,
    \quad
    B^T \mathbb{W}\sigma_0 = f .
\end{equation}
We verify that the dimension of $Z$ is $2N = 2m$ and the dimension of $E$ is $N = m = n + l$. We may regard $(u,v) \in \mathbb{R}^n \times \mathbb{R}^l$ as a set of coordinates parameterizing $E$. In this representation, the posterior likelihood function takes the form
\begin{equation}\label{nx9SKf}
    L(u,v)
    =
    L_D(Bu+\epsilon_0, \mathbb{W}^{-1}Av+\sigma_0) ,
\end{equation}
where we write $L_D(\epsilon,\sigma)$ for the material likelihood function. The corresponding posterior probability functions then follows by normalization of $L(u,v)$. For instance, if $L_D$ is of the form (\ref{0VUrvV}), representation (\ref{nx9SKf}) gives
\begin{equation}
    L(u,v)
    =
    \exp
    \Big(
        -
        \frac{1}{2s^2}
        \Big(
            \| Bu+\epsilon_0 \|^2 + \| \mathbb{W}^{-1}Av+\sigma_0 \|^2 - 2 (\mathbb{W}^{-1}Av+\sigma_0) \cdot (Bu+\epsilon_0)
        \Big)
    \Big) ,
\end{equation}
or, using the orthogonality properties of $A$ and $B$,
\begin{equation}
    L(u,v)
    =
    \exp
    \Big(
        -
        \frac{1}{2s^2}
        \Big(
            \| Bu+\epsilon_0 \|^2
            +
            \| \mathbb{W}^{-1}Av+\sigma_0 \|^2
            -
            2 \mathbb{W}^{-1}Av \cdot \epsilon_0
            -
            2 \sigma_0 \cdot Bu
        \Big)
    \Big) .
\end{equation}
From the properties of Gaussian integrals, the expected values $(\bar{u},\bar{v})$ of the coordinates $(u,v)$ are obtained by minimizing the potential $\Phi=-\log(L)$, with the result,
\begin{subequations}
\begin{align}
    &
    \bar{u} = (B^T\mathbb{W}\mathbb{C}B)^{-1} B^T\mathbb{W}(\sigma_0 - \mathbb{C} \epsilon_0) ,
    \\ &
    \bar{v} = (A^T\mathbb{W}^{-1}\mathbb{C}^{-1}A)^{-1} A^T(\epsilon_0 - \mathbb{C}^{-1} \sigma_0) ,
\end{align}
\end{subequations}
which are computed by inverting the stiffness and compliance matrices of the truss, $B^T \mathbb{W}\mathbb{C} B$ and $A^T\mathbb{W}^{-1}\mathbb{C}^{-1}A$, respectively, with $\mathbb{C}={\rm diag}(\mathbb{C}_1, \cdots, \mathbb{C}_m)$.
The posterior probability density then follows as
\begin{equation}
    {L}(u,v)
    =
    \sqrt{\det(B^T \mathbb{W}\mathbb{C} B /2\pi s^2 )}
    \sqrt{\det(A^T \mathbb{W}^{-1}\mathbb{C}^{-1} A /2\pi s^2 )}
    \times
\end{equation}
\begin{equation*}
    \exp
    \Big(
        -
        \frac{1}{2s^2}
        \Big(
            (B^T\mathbb{W}\mathbb{C}B)(u-\bar{u}) \cdot (u-\bar{u})
            +
            (A^T\mathbb{W}^{-1}\mathbb{C}^{-1}A)(v-\bar{v}) \cdot (v-\bar{v})
        \Big)
    \Big) ,
\end{equation*}
which provides a full account of the probability of outcomes. \hfill$\square$
\end{example}

\subsection{Approximation by empirical data} \label{8yK9TU}

Suppose, contrariwise, that, as in commonly the case in practice, the material likelihood function $L_D$ is known only approximately through a sequence of empirical samples, e.~g., drawn from experimental measurements, consisting of point-data sets $P_h = \{ y_{h,i},\ i=1,\dots,M_h \} \subset Z$, $h=0,1,\dots$, of size $M_h$, cf.~Fig.~\ref{uCXq6P}b. We additionally associate with each discrete data point a certain likelihood $c_{h,i} \in [0,1]$ accounting, e.~g., for experimental reliability and scatter \cite{kirchdoerfer2017data}. Thus, $c_{h,i} = 1$ for a data point $y_{h,i}$ that is sure to be free of experimental error, whereas $c_{h,i} < 1$ if the experimental measurements show scatter or are of questionable provenance. For every material point-set $P_h$, we consider discrete expectations of the general form


\begin{equation}\label{a2VR0G}
    \mathbb{E}_h[f]
    =
    \sum_{i=1}^{M_h}
    \Big( \int_E p_{h,i}(z) f(z) \, dz \Big),
\end{equation}


where $f \in C_b(Z)$ is a continuous function representing a quantity of interest,
and we require


\begin{equation}\label{fREO9W}
    p_{h,i}(z) \geq 0 ,
    \quad
    \sum_{i=1}^{M_h}
    \Big( \int_E p_{h,i}(z) \, dz \Big)
    =
    1 .
\end{equation}


We envision an experimental campaign resulting in point-data samples $P_h$ of increasing size, $M_h \uparrow \infty$ and increasing fidelity $c_{h,i} \uparrow 1$. The central question is, then, under what conditions on the data, if any, and what choices of probabilities $p_{h,i}(z)$ the approximate expectations $\mathbb{E}_h[f]$ {\sl converge} to the exact ones $\mathbb{E}[f]$ in the limit as $h\to\infty$.

We specifically consider probabilities $p_h \equiv \{p_{h,i}(z),\ i=1,\dots,M_h\}$ that minimize the regularized Kullback-Leibler (KL) divergence \cite{Kullback:1951, Kullback:1959}


\begin{equation}\label{Qg3xQI}
    G_\beta(p_h)
    =
    \sum_{i=1}^{M_h}
    \Big(
        \int_E
            \Big(
                \beta \| y_{h,i} - z \|^2
                +
                \log\dfrac{p_{h,i}(z)}{c_{h,i}}
                -
                N \log\frac{\beta}{\beta_0}
            \Big)
            \, p_{h,i}(z)
        \, dz
    \Big) ,
\end{equation}
for some reference $\beta_0 > 0$. We note that the form (\ref{a2VR0G}) assumed of the trial expectations ensures that the implied trial likelihoods are absolutely continuous with respect to the assumed material point-data set and deterministic loading, a requirement of the KL divergence. Evidently, the divergence function (\ref{Qg3xQI}) penalizes distance to the constraint set $E$, i.~e., it assigns an increasing discrepancy to material data points $y_{h,i}$ that are far from $E$, while simultaneously attempting to match the {\sl a priori} likelihood $c_{h,i}$ of the point. The additional term $N \log(\beta/\beta_0)$ in (\ref{Qg3xQI}) is introduced in order to ensure that the KL functions $(G_\beta)$ are uniformly bounded with respect to $\beta$ and it only plays a role in the variational characterization of $\beta$, cf.~Section~\ref{6BugTf}. 

The optimal regularized discrete probability $p_h$ then follows by minimization of $G_\beta$, with the explicit result
\begin{equation}\label{zNR2jN}
    p_{h,i}(z)
    =
    \frac
    {
        c_{h,i} 
        {\rm e}^{-\beta\|y_{h,i}-z\|^2}
    }
    {
        \sum_{i=1}^{M_h} \int_E c_{h,i} 
        {\rm e}^{-\beta\|y_{h,i}-z\|^2} \, dz
    } ,
    \qquad
    z \in E .
\end{equation}
We see from this expression that material data points $y_{h,i}$ that are far from the constraint set $E$ contribute less than points are are close, as expected. The parameter $\beta$ sets the scale of the distance as $1/\sqrt{\beta}$.

\subsection{Annealing schedule}
\label{6BugTf}

We note that, for $\beta \to +\infty$, the posterior probabilities (\ref{zNR2jN}) degenerate in general. This degeneracy simply reflects the fact that the point sets $P_h$ are unlikely to have a consequential intersection with the admissible set $E$, i.~e., the likelihood that a point $y_{h,i}$ be admissible with respect to a given deterministic loading is zero. As $\beta \to 0$, the posterior probabilities (\ref{zNR2jN}) also degenerate. There must therefore be an optimal value of $\beta$ where the probabilities are least degenerate. Conveniently, this optimal value $\beta_h$ of $\beta$ again follows by minimization of the KL divergence (\ref{Qg3xQI}). Thus, evaluating $G_\beta(p_h)$ at the minimizer (\ref{zNR2jN}), we obtain


\begin{equation}
    \min G_\beta
    =
    -
    \log
    \Big(
        \sum_{i=1}^{M_h} \int_E c_{h,i}
        {\rm e}^{-\beta\|y_{h,i}-z\|^2} \, dz
    \Big)
    -
    N \log \frac{\beta}{\beta_0} .
\end{equation}
Minimizing with respect to $\beta$ gives the optimality condition
\begin{equation}\label{cDTXAV}
    \frac{1}{\beta_h}
    =
    \frac{1}{N}
    \frac
    {
        \sum_{i=1}^{M_h}
        \int_E
            c_{h,i}
            \|y_{h,i}-z\|^2
            {\rm e}^{-\beta_h \|y_{h,i}-z\|^2}
        \, dz
    }
    {
        \sum_{i=1}^{M_h}
        \int_E
            c_{h,i}
            {\rm e}^{-\beta_h \|y_{h,i}-z\|^2}
        \, dz
    } ,
\end{equation}
which defines $\beta_h$ implicitly. Finally, inserting (\ref{cDTXAV}) into (\ref{a2VR0G}) we obtain
\begin{equation}\label{Pa55uo}
    \mathbb{E}_h[f]
    =
    \frac
    {
        \sum_{i=1}^{M_h}
        \Big(
            \int_E
                c_{h,i}
                {\rm e}^{-\beta_h \|y_{h,i}-z\|^2}
                f(z)
            \, dz
        \Big)
    }
    {
        \sum_{i=1}^{M_h}
        \Big(
            \int_E
                c_{h,i}
                {\rm e}^{-\beta_h \|y_{h,i}-z\|^2}
            \, dz
        \Big)
    } ,
\end{equation}


which supplies an expression for the approximate expectations that is explicit up to quadratures.

\subsection{Factorization and localization}

By exchanging the integrals over $E$ and the sums over data in (\ref{Pa55uo}), the approximate expectations have the equivalent form
\begin{equation}\label{tQ6bZL}
\begin{split}
    \mathbb{E}_h[f]
    =
    \frac
    {
        \int_E
        \sum_{i=1}^{M_h}
            f(z)
            {\rm e}^{-\beta_h \|y_{h,i}-z\|^2}
            c_{h,i}
        \, dz
    }
    {
        \int_E
        \sum_{i=1}^{M_h}
            {\rm e}^{-\beta_h \|y_{h,i}-z\|^2}
            c_{h,i}
        \, dz
    } .
\end{split}
\end{equation}
The great advantage of this alternative form is that it factorizes with respect to local data, which eliminates the combinatorial complexity of the first form. Thus, suppose that
\begin{equation}
    D_h = \prod_{e=1}^m D_{e,h} ,
\end{equation}
where $D_{e,h} = \{ y_{e,h,i},\ i=1,\dots,M_{e,h} \}$  are local data files. Then,
\begin{equation}\label{0lgoUa}
    \int_E
    \sum_{i=1}^{M_h}
        f(z)
        {\rm e}^{-\beta_h \|y_{h,i}-z\|^2}
        c_{h,i}
    \, dz
    =
    \int_E
        f(z) {L}_h(z)
    \, dz ,
\end{equation}
where
\begin{equation}
    {L}_h(z)
    =
    \sum_{i=1}^{M_h}
        {\rm e}^{-\beta_h \|y_{h,i}-z\|^2}
        c_{h,i}
\end{equation}
is an inferred likelihood density. For local data, the sum over data factorizes, with the result
\begin{equation}
    {L}_h(z)
    =
    \prod_{e=1}^m {L}_{e,h}(z_e) ,
\end{equation}
where
\begin{equation}\label{6r8WgM}
    {L}_{e,h}(z_e)
    =
    \sum_{i=1}^{M_{e,h}}
        {\rm e}^{-\beta_h \|y_{e,h,i}-z_e\|^2}
        c_{e,h,i}
\end{equation}
are local likelihood density functions. We note that the computational complexity of the expectation is $\sum_{e=1}^m M_{e,h}$, instead of $\prod_{e=1}^m M_{e,h}$, as before, which renders the evaluation of approximate expectations computationally feasible.

We further note that, for local data, we can generalize (\ref{6r8WgM}) to
\begin{equation}
    {L}_{e,h}(z_e)
    =
    \sum_{i=1}^{M_{e,h}}
        {\rm e}^{-\beta_{e,h} \|y_{e,h,i}-z_e\|^2}
        c_{e,h,i} ,
\end{equation}
where we localize temperature. We can then estimate $1/\beta_{e,h}$ as the average minimum distance squared between points in $D_{e,h}$, i.~e.,
\begin{equation}\label{eq:beta_avg}
    \frac{1}{\beta_{e,h}}
    \sim
    \frac{1}{M_{e,h}}
    \sum_{i=1}^{M_{e,h}}
    \min_{j \neq i} \| y_{e,h,i} - y_{e,h,j} \|^2 .
\end{equation}

This local approximation of Eq. (\ref{cDTXAV}) has the advantage that $\beta_{e,h}$ can be explicitly pre-computed for each local data set and is therefore used in the following numerical examples.

Finally, the integrals over $E$ can be conveniently computed by stochastic quadrature. These and other matters of implementation are discussed next.

\section{Numerical implementation}

\subsection{Problem statement}
In evaluating (\ref{tQ6bZL}), two main operations need to be performed: the integration over the constraint set $E$ and the summation over the global material data set for the evaluation of the material likelihood. The main computational challenges pertaining to these operations are:
\begin{enumerate}
    \item The evaluation of the local material likelihood involves a summation over local data sets for each material point that can be computationally demanding. However, most data points have negligible influence on the total sum, which suggests the use of efficient, carefully designed {\sl a propos} data structures.
    \item The Monte Carlo integration over the constraint set $E$ requires the generation of random points on $E$, which requires the repeated solution of global systems of equations if the sampling is done by projection. It is therefore advantageous to identify sampling strategies that reduce the number of projections.
    \item The thermalized likelihood function implied by (\ref{Pa55uo}) is exceedingly multi-modal, featuring peaks at every data point (cf.~\cite{kanno2019mixed} for a discussion in the context of {\sl min-dist} DD). This intricate likelihood landscape arises even if the underlying likelihood function is smooth and compounds the evaluation of the integrals over $E$ in (\ref{Pa55uo}).
\end{enumerate}
These challenges are addressed next in turn.

\subsection{Population annealing}
The complexity of the thermalized likelihood landscape renders the direct application of standard Markov chain importance sampling strategies, such as the classical Metropolis--Hastings algorithm \cite{metropolis1953equation}, difficult. Therefore, a more tailored sampling strategy is required, as discussed next.

\subsubsection{Population Annealing}
{\sl Population Annealing} (PA) \cite{iba2001population, machta2010population, weigel2021understanding} is a sequential Monte Carlo method designed for complex likelihood functions and scalable parallel computations. A set of states, the population, is maintained in order to explore different regions of phase space that are possibly separated by energy barriers resulting from local minima. Each population member performs a relatively short Markov chain importance sampling. In addition, the population is resampled by a combination of pruning and enriching, in analogy to {\sl Genetic Algorithms}. The process is driven by a {\sl Simulated Annealing} schedule, which slowly increases the inverse local temperature $\beta_e$. For every new temperature, the population is re-equilibrated by resampling and a specified number of Markov chain moves. In this manner, at low $\beta_e$ the population can escape readily from local minima and perform a more exhaustive exploration. At high $\beta_e$, the population is increasingly driven towards the constraint set $E$.

\subsubsection{Initialization and resampling step}
The principle idea of the resampling step is to remove and duplicate the population samples according to their energy while increasing $\beta_e$ in order to accelerate the convergence to low-energy states. Resampling is done in such a way that the population size fluctuates around a target population size $N_P^*$ and equilibrium of the distribution is maintained for every new temperature.

Initially, at $\beta_e=0$, a population $\mathcal{P} = \{z_p\in E \, : \, p=1\dots N_{P}\}$ of $N_{P}$ states is generated either randomly by the closest-point projection of random global data points onto $E$ or using a {\sl min-dist} DD solver \cite{kirchdoerfer2016data}. At initialization, the population size $N_P$ may be chosen equal to the target size $N_P^*$, or set to a different value, in which case the size gets adapted to the target size during resampling. The energy of the population is initially set to infinity. For $\beta_e>0$, the energy is defined as
\begin{equation}\label{eq:energy}
    e_p
    =
    -
    \frac{1}{\beta_e} \,
    \sum_{e=1}^{m}
        \log
        \Big(
            \frac{1}{M_e}
            \sum_{i=1}^{M_e}
            {\rm e}^{-\beta_e\|y_{e,i}-z_{e,p}\|^2}
        \Big).
\end{equation}
where we assume local data and $c_i=1$ for all data points. The efficient evaluation of (\ref{eq:energy}) is discussed in Section \ref{sec:energy}.

At each iteration, or {\sl quench}, $\beta_e$ is increased according to
\begin{equation}
    \beta_e = \beta_e+\Delta\beta_e
\end{equation}
for a specified number of quenches $N_Q$. For simplicity, in calculations we choose a constant increment
\begin{equation}
    \Delta\beta_e = \beta_{e,f}/N_Q
\end{equation}
for a pre-computed target $\beta_{e,f}$. The simple choice of $\beta_{e,f}$ given by the inverse average nearest neighbor distance (\ref{eq:beta_avg}) is used in the numerical examples presented in Sections~\ref{sec:three_bars_Gaussian} to \ref{sec:pactruss_study}.

For resampling, the relative weights
\begin{equation}
    \hat\tau_p = N_P^*\frac{{\rm e}^{-\Delta\beta_e e_p}}{\sum_{i=1}^{N_{P}}{\rm e}^{-\Delta\beta_e e_i}}
\end{equation}
are computed for each population member. The required number of copies of population member $p$ is then conveniently estimated as \cite{wang2015comparing,barash2017gpu}
\begin{align}
    l_p
    =
    \begin{cases}
        \lfloor\hat\tau_p\rfloor, &  \text{if}~ u>\hat\tau_p-\lfloor\hat\tau_p\rfloor ,\\
        \lfloor\hat\tau_p\rfloor+1, & \text{otherwise} ,
    \end{cases}
\end{align}
where $u$ is a random number drawn uniformly from $[0,1)$ and the floor function $\lfloor x\rfloor$ is the largest integer less or equal $x$. If $l_p=0$, member $p$ is removed from the population. The resampling step incurs a modest computational overhead that is offset by the overall performance gain resulting from reconfiguring the population to the relevant regions of the phase space.

\subsubsection{Markov chain importance sampling step}\label{sec:pca}
Resampling introduces a bias since population members may be duplicated. In order to eliminate this bias, a Markov chain importance sampling step is performed. In this step, every population member executes a random walk, thereby exploring new states and decorrelating the population. The random walk consists of repeated random moves in the constraint set $E$. We recall that the constraint set $E$ is a linear subspace of dimension $N=md$ (number of material points times dimension of stress/strain) in an $2N-$dimensional global phase space $Z$. We can conveniently exploit this affine structure and represent $E$ in terms of an $N$-dimensional basis.

We compute the requisite basis by means of a principal component analysis (PCA). To this end, we generate $K>N$ random global data points $z_i \in Z$ and project them onto $E$. For consistency with the norm (\ref{Poyet2}), we further weight the strains and stresses by $\sqrt{w_e}\mathbb{C}_e^{1/2}$ and $\sqrt{w_e}\mathbb{C}_e^{-1/2}$, respectively. The weighted states $z_i^w$ (here and subsequently the superscript $w$ indicates weighting) are collected in a matrix $T\in\mathbb{R}^{K\times 2N}$ in order to compute its covariance matrix $C_E\in\mathbb{R}^{2N\times 2N}$. Computing the eigenvectors of $C_E$ associated to the $N$ largest eigenvalues then gives the basis vectors of the constraint set as a matrix $A_E\in\mathbb{R}^{2N\times N}$.

To effect a move, $N$ random numbers are drawn from a standard normal distribution and multiplied by a step size $s_p$, transformed back by $A_E^T$ and added to the weighted state $z^w_p$
\begin{equation}
    z^w_{p,\rm trial} = z^w_p + s_p~ g~A_E^T, \quad g\sim\mathcal{N}(0,I_N) .
\end{equation}
The proposed move to the trial state $z_{p,\rm trial}$ (obtained by reverse weighting of $z^w_{p,\rm trial}$) is then either accepted or rejected according to the Metropolis-Hastings acceptance probability
\begin{equation}
    P_{\beta_e}=\min(1,\exp(-\beta_e(e_{p,\rm trial}-e_p))) ,
\end{equation}
where $e_{p,\rm trial}$ is the energy of $z_{p,\rm trial}$. This acceptance criterion allows moves that increase the energy slightly, in particular at the initial stages of annealing characterized by low $\beta_e$. In addition, the step size $s_p$ influences the acceptance rate and, therefore, the efficiency of the algorithm. In calculations, we chose the simple adaption scheme
\begin{equation}
    s_p = s_p + (r_p-r^*)s_p ,
\end{equation}
with a target acceptance rate $r^*\in[0,1]$. The acceptance rate $r_p$ is measured as the ratio of accepted moves to the total number of trials $N_T$. The adaption scheme has the effect of reducing the step size if the acceptance rate is smaller than $r^*$, which signals an overly-aggressive exploration strategy, and increasing it if the acceptance rate is too small, which contrariwise signals an overly-timid exploration strategy.

\subsubsection{Energy computation}\label{sec:energy}
The complexity of the evaluation of the energy (\ref{eq:energy}) is $\sum_e M_e$. However,  for sufficiently large $\beta_e$ a vast majority of the data points contribute negligibly to(\ref{eq:energy}), which suggests restricting the sum with the aid of supporting data structures. In \cite{eggersmann2021efficient}, a number of data structures for local data sets were investigated in the framework of min-dist DD solvers and the hierarchical k-means tree \cite{fukunaga1975branch} was found to be particular well-suited for executing approximate nearest neighbor searches.

In this work, we likewise use the hierarchical $k$-means implementation of the FLANN library \cite{muja2014scalable}. The approximate nearest neighbor search provided by the library is performed using a priority queue of maximum length $N_{\rm checks}$ in order to reduce backtracking operations. We employ a radius search technique, which returns all data points $y_{e,i}\in D_e$ within a specified radius
\begin{equation}
    r^2_{\rm TOL}=-\frac{1}{\beta_e}\log({\rm TOL}).
\end{equation}
around the query point $z_{p,e}$. This technique has the advantage over a $k$-nearest-neighbor ($k$NN) search that the number of nearest neighbors $k$ need not be specified. Instead, the sum is restricted to points whose contribution exceeds the tolerance ${\rm TOL}$.

Our numerical tests reveal that using a radius search increases performance markedly and that, remarkably, the performance gains are insensitive to the choice of tolerance. Thus, in tests values of ${\rm TOL} = $  $10^{-8},10^{-16},10^{-32}$ delivered nearly identical performance. In view of this tradeoff, in all calculations presented subsequently we use ${\rm TOL}=10^{-16}$. By contrast, the performance of the algorithm is strongly dependent on the degree of backtracking as set by $N_{\rm checks}$, cf.~Section \ref{sec:pactruss_study}.

It bears emphasis that only one material data structure needs to be set up for each specific material in the structure, since all members made of the same material access the same local data set. Likewise, only one inverse local temperature $\beta_{e,f}$ and one quenching scheme is required for each specific material. In particular, if all structural members are made of the same material, only one material data structure and quenching scheme need to be maintained and run.

\subsubsection{Summary of the algorithm}
The described algorithm is summarized in Algorithm \ref{alg:PA}. It returns the final population $\mathcal{P}$ of size $N_P$. The expectation of a quantity of interest $f$ is evaluated as
\begin{equation}
    \mathbb{E}[f]=\frac{1}{N_{P}}\sum_{p=1}^{N_{P}} f(z_p), \quad z_p\in \mathcal{P}.
\end{equation}
One advantage of the algorithm is that does not require extensive fine-tuning of the input variables. Thus, the target population size $N_{P}^*$, the number of quenches $N_Q$ and the number of Markov chain moves $N_T$ are parameters that are chosen as large as permitted by computational resources. In all calculations, the target acceptance rate is set to $r^*=0.25$. The initial step size $s_{0}$ influences only the first few iterations as it is adapted during the quenching schedule. The step size tends to decrease with increasing dimensionality \cite{graves2011automatic}.
Lastly, we note that the population member perform their moves independently of each other, which renders the algorithm {\sl embarrassingly parallel}.

\begin{algorithm}[H]
\caption{Population annealing algorithm}
\label{alg:PA}
\begin{algorithmic}
\REQUIRE Constraint set $E \subset Z$, basis vectors $A_E$.
\REQUIRE Local data set(s) $D_e$
\REQUIRE $N_Q,N_T,N_{P}^*,s_{0},r^*,\beta_{e,f}$
\STATE Initialize: $\mathcal{P}\leftarrow\{z_p\in E\}$ randomly or heuristically
\STATE Initialize: step size $s_p\leftarrow s_{0}$
\STATE Initialize: energy $e_p\leftarrow \infty$
\STATE Initialize: $\beta_e\leftarrow0$
\STATE Set: $\Delta\beta_e\leftarrow\beta_{e,f}/N_Q$

\FOR{$q=0,\dots,N_Q-1$}
\STATE Set: $\beta_e\leftarrow\beta_e+\Delta\beta_e$
\STATE Set: $\mathcal{P}\leftarrow\texttt{Resampling}(\mathcal{P},\Delta\beta_e)$

\FOR{\textbf{each} $z_p\in \mathcal{P}$}
\FOR{$t=0,\dots,N_T-1$}
\STATE Set: $z_{p,\rm trial}\leftarrow \texttt{RandomMove}(z_p,s_p,A_E)$
\STATE Set: $e_{p,\rm trial}\leftarrow \texttt{Energy}(D_e,\beta_e,z_{p,\rm trial})$

\STATE Set: $u\leftarrow \texttt{UniformRand}(0,1)$
\STATE Set: $w \leftarrow \exp(- \beta_e(e_{p,\rm trial} -e_p))$
\IF {$u\leq w$}

\STATE Set: $z_p\leftarrow z_{p,\rm trial}$
\STATE Set: $e_p\leftarrow e_{p,\rm trial}$
\ENDIF
\ENDFOR
\ENDFOR
\STATE Set: $r_p\leftarrow\texttt{AcceptanceRate}$
\STATE Set: $s_p\leftarrow s_p+(r_p-r^*)s_p$
\ENDFOR

\RETURN $\mathcal{P}$
\end{algorithmic}
\pagebreak
\end{algorithm}

\section{Examples of application}

We conclude with selected examples of application that illustrate simply the range, scope and convergence properties of the approach.

\subsection{Three-bar truss with sliding-Gaussian material data set}\label{sec:three_bars_Gaussian}

\begin{figure}[ht]
	\begin{subfigure}{0.45\textwidth}\caption{} \includegraphics[width=0.99\linewidth]{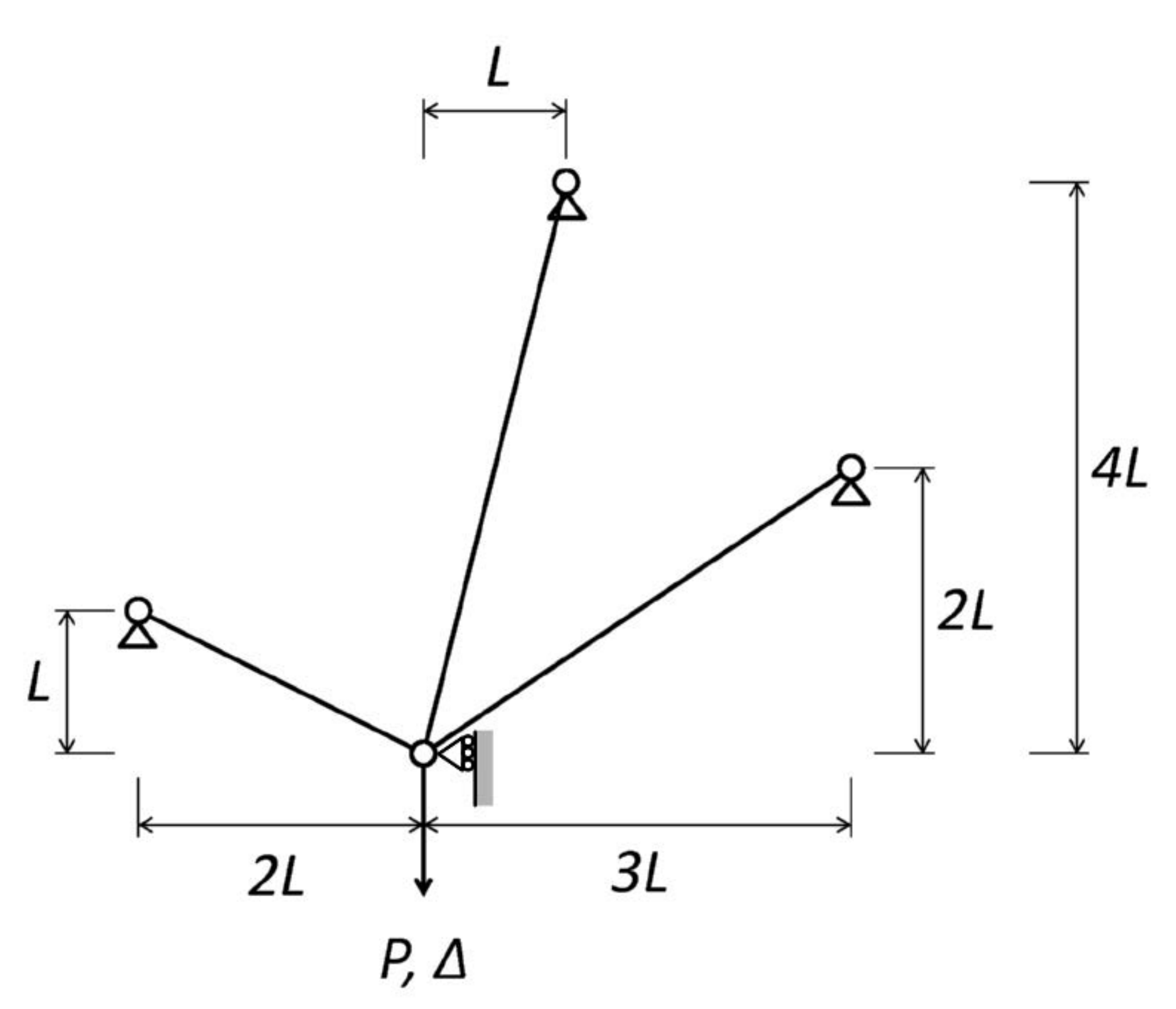}
	\end{subfigure}
    $\quad$
	\begin{subfigure}{0.40\textwidth}\caption{} \includegraphics[width=0.99\linewidth]{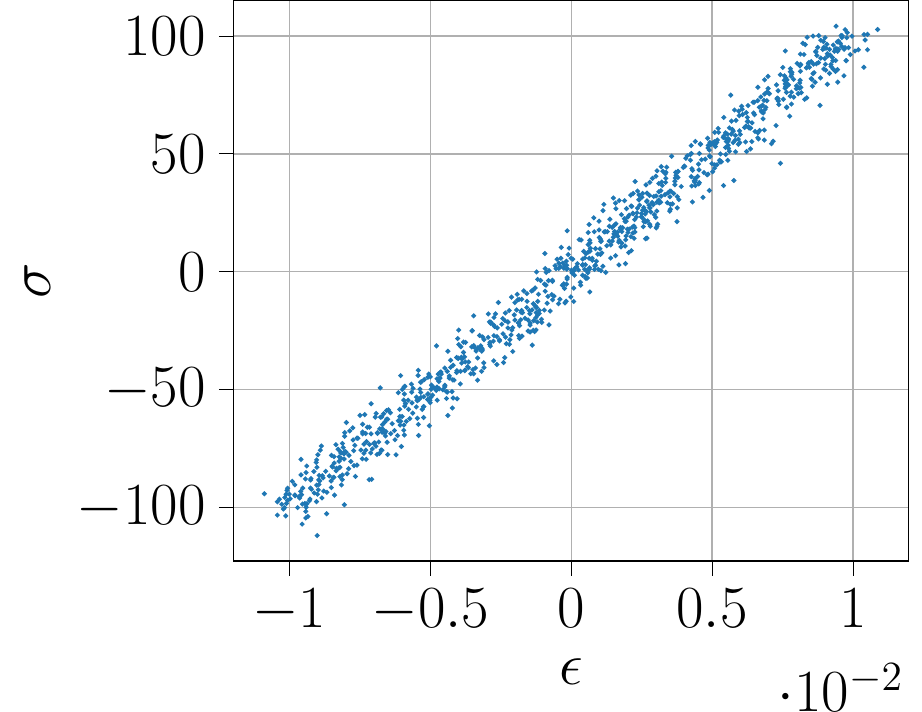}
	\end{subfigure}
    \caption{a) Truss with prescribed applied force $P$ undergoing a displacement $\Delta$. b) Local material data set with $1000$ data points sampled from a sliding-Gaussian likelihood function.} \label{fig:truss_gauss}
\end{figure}

We begin with a simple verification example concerned with the three-bar truss shown in Figure \ref{fig:truss_gauss}a loaded by a force $P$ and undergoing a corresponding displacement $\Delta$. In calculations, we set $P=100$, $L=1$ and all cross-sectional areas $A=1$ (here and subsequently units are omitted). The outcome quantity of interest is chosen to be the displacement $\Delta$.

In order to have an analytical reference solution, we assume that the material data set obeys a sliding-Gaussian distribution, in which case the exact solution of the inference problem is given in Section~\ref{suO4Ci}. The local material data set is generated according to Section \ref{dzeXXX} with $\mathbb{C}_e=10000$ and standard deviation $s=5 \times 10^{-4}$. A typical data set is shown in Fig.~\ref{fig:truss_gauss}b. We emphasize that such material data sets are the only input to the calculations and, once generated, the underlying distribution whence they are sampled is jettisoned altogether and presumed unknown.

Calculations are carried out using Algorithm~\ref{alg:PA} with number of quenches $N_Q=100$, number of trials $N_T=20$ and initial step size $s_{0}=1.0$. Nearest-neighbor searches with radius ${\rm TOL}=10^{-16}$ are performed exactly, i.~e., with an unlimited number of backtracking checks $N_{\rm checks}$. As a measure of the error we choose the Kolmogorov–Smirnov statistic $e_{\rm KS}$, which measures the maximum difference between the exact and computed cumulative distributions.

\begin{figure}[ht]
	\begin{subfigure}{0.30\textwidth}\caption{} \includegraphics[width=0.99\linewidth]{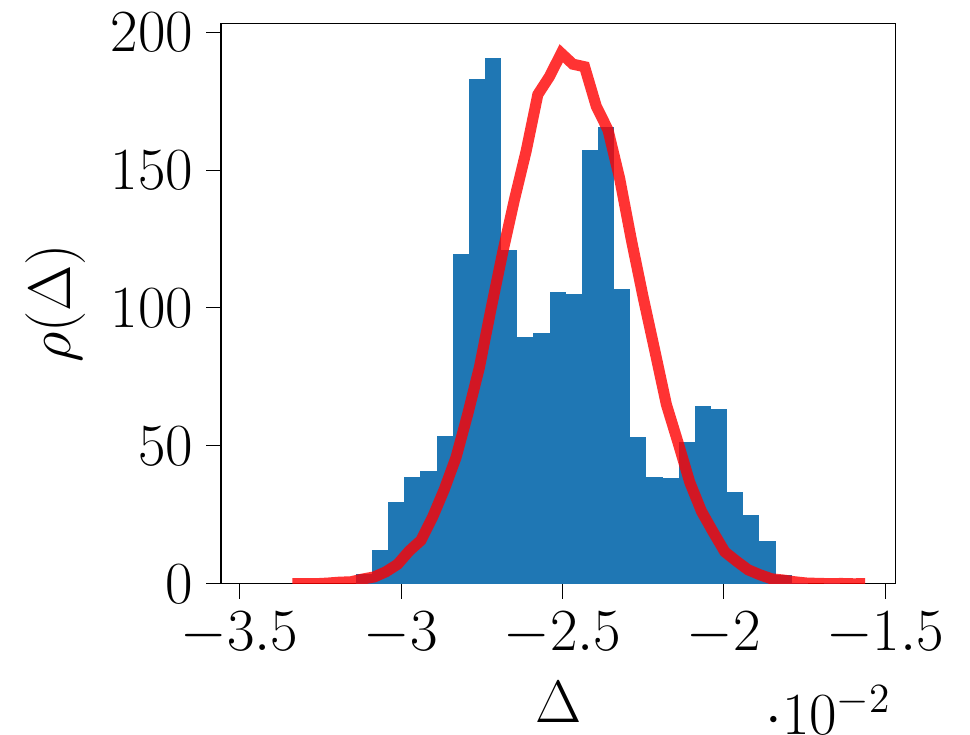}
	\end{subfigure}
	\begin{subfigure}{0.30\textwidth}\caption{} \includegraphics[width=0.99\linewidth]{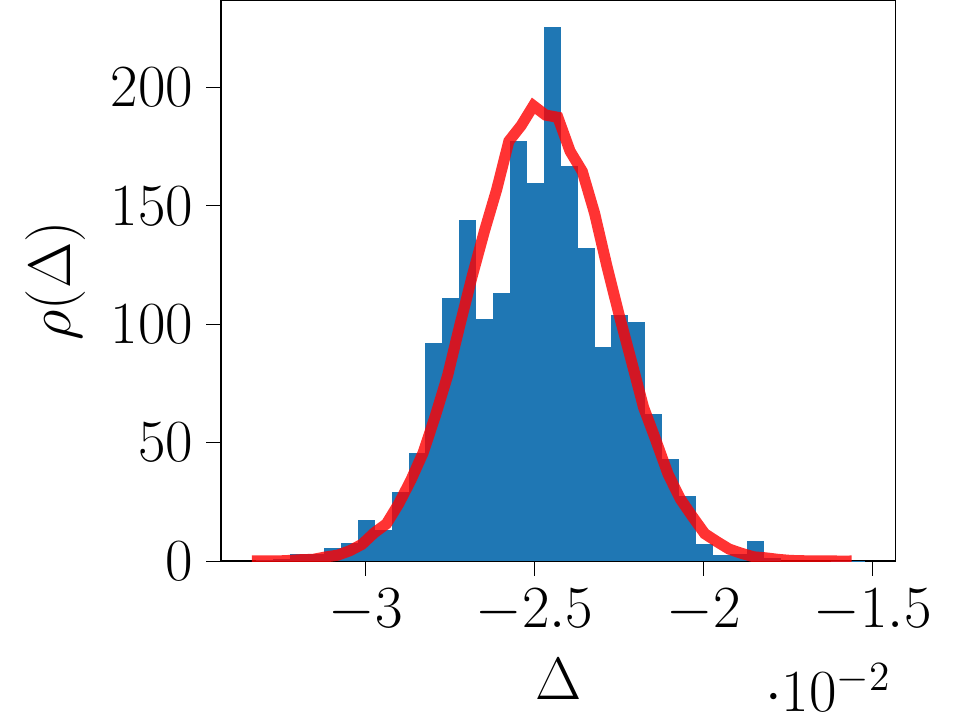}
	\end{subfigure}
    	\begin{subfigure}{0.30\textwidth}\caption{} \includegraphics[width=0.99\linewidth]{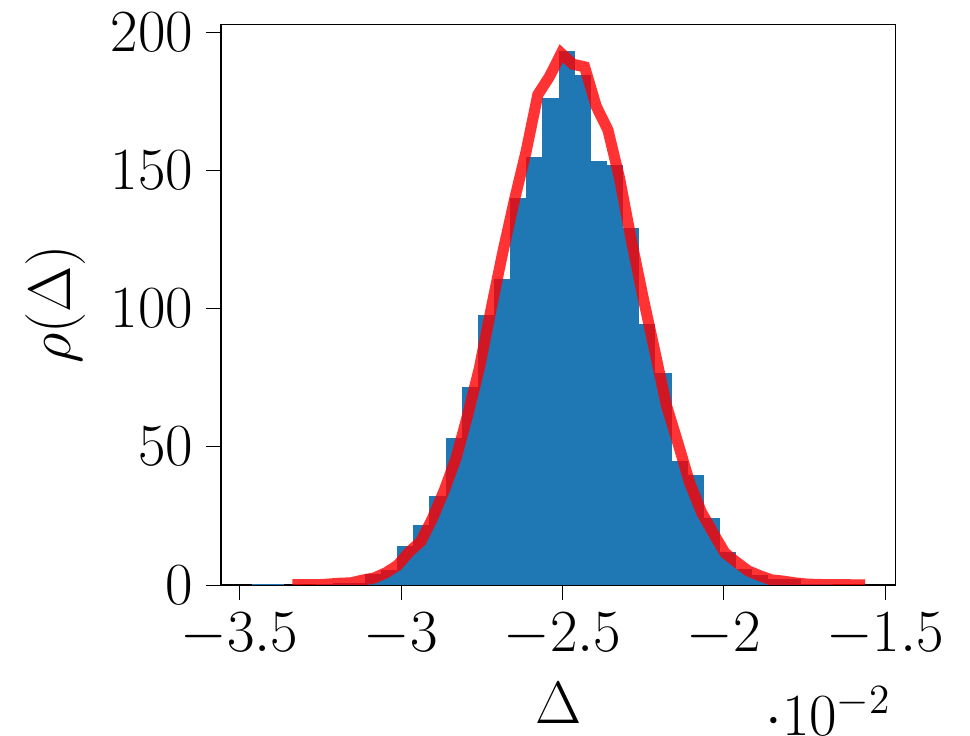}
	\end{subfigure}
    \caption{Three-bar truss with sliding-Gaussian material data. Computed histograms {\sl vs.} exact distribution (red) of displacement $\Delta$ for material data-sets of sizes: a) $M_e=10^3$; b) $M_e=10^4$; c) $M_e=10^5$.} \label{fig:gauss_pdf_study}
\end{figure}

Fig.~\ref{fig:gauss_pdf_study} collects three computed histograms of displacement $\Delta$ for three different material data-set sizes $\{10^3,10^4,10^5\}$. The exact Gaussian distribution is shown in red for comparison. A clear trend towards convergence of the computed histogram to the exact distribution is apparent from the figure. It bears emphasis that, as expected, both the exact distribution and the computed histograms of displacement are strictly {\sl unimodal}. Unimodality is indicative of a lack of complexity of the behavior of a system. In particular, the maximum-likelihood state of the system---equivalently, the minimum free-energy solution---is uniquely defined \cite{kirchdoerfer2017data} and the outcomes of the system cluster around the maximum-likelihood state.
For the given problem, the maximum-likelihood solution is computed to $\Delta_{\rm ML}=-2.48\cdot 10^{-2}$. 

\begin{figure}[ht]
	\begin{subfigure}{0.45\textwidth}\caption{}
	\centering
	\includegraphics[height=5cm]{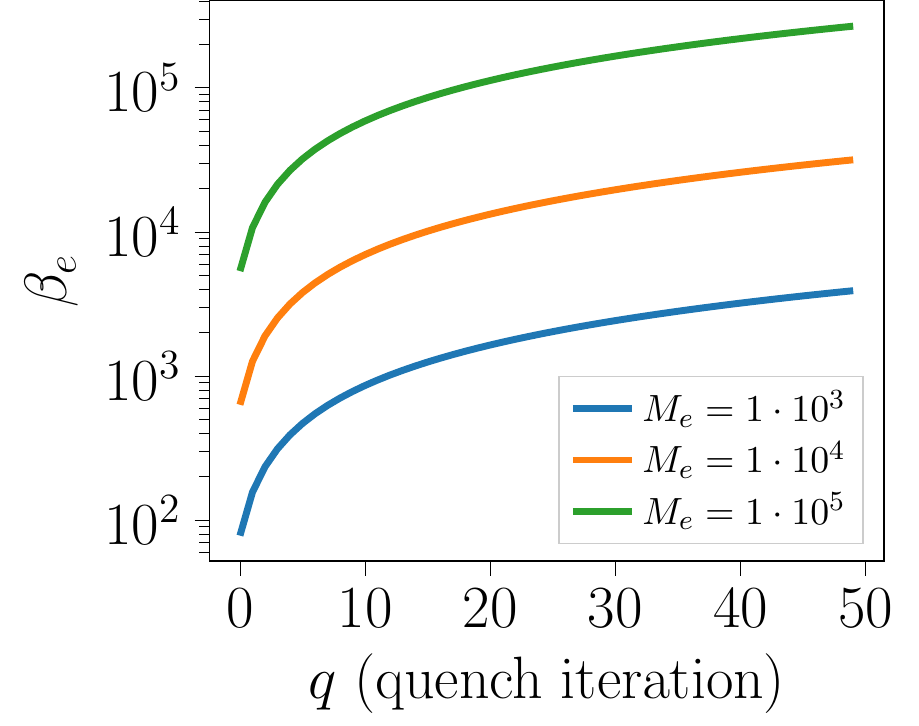}
	\end{subfigure}
	\begin{subfigure}{0.45\textwidth}\caption{}
	\centering
	\includegraphics[height=5cm]{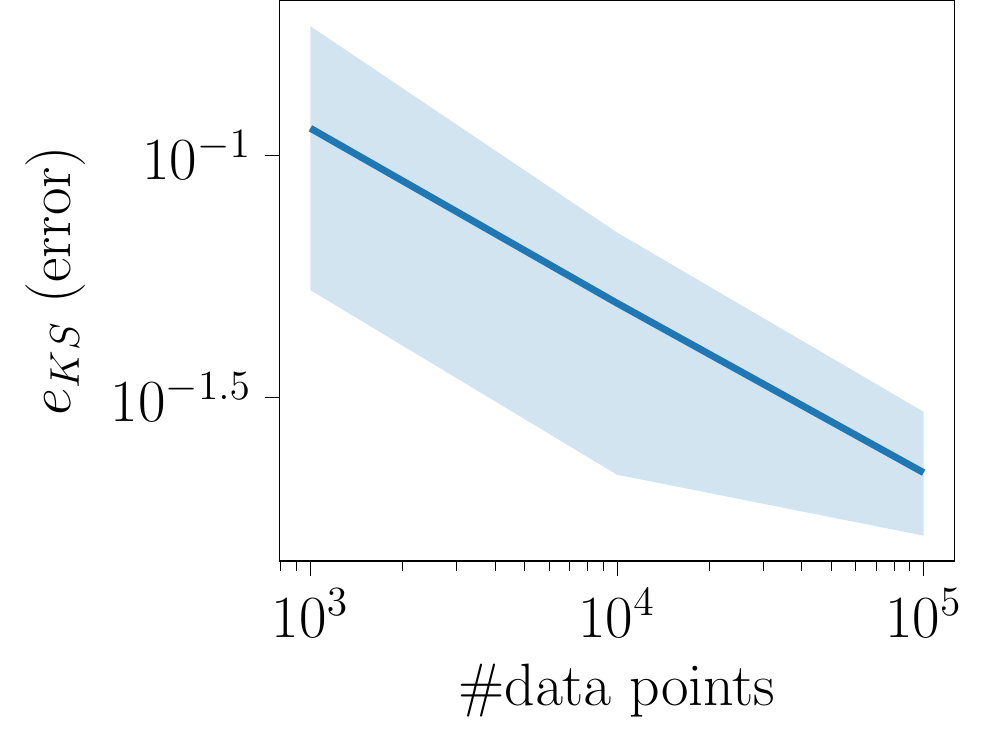}
	\end{subfigure}
    \caption {Three-bar truss with sliding-Gaussian material data. Convergence of temperatures and computed histograms of displacement $\Delta$. a) Evolution of $\beta_e$ during annealing process. b) Kolmogorov–Smirnov (KS) error {\sl vs.} material data-set size, $N_P=10^4$.
    } \label{fig:data_set_error}
\end{figure}

\begin{figure}[ht]
	\begin{subfigure}{0.45\textwidth}\caption{}
	\centering
	\includegraphics[height=5cm]{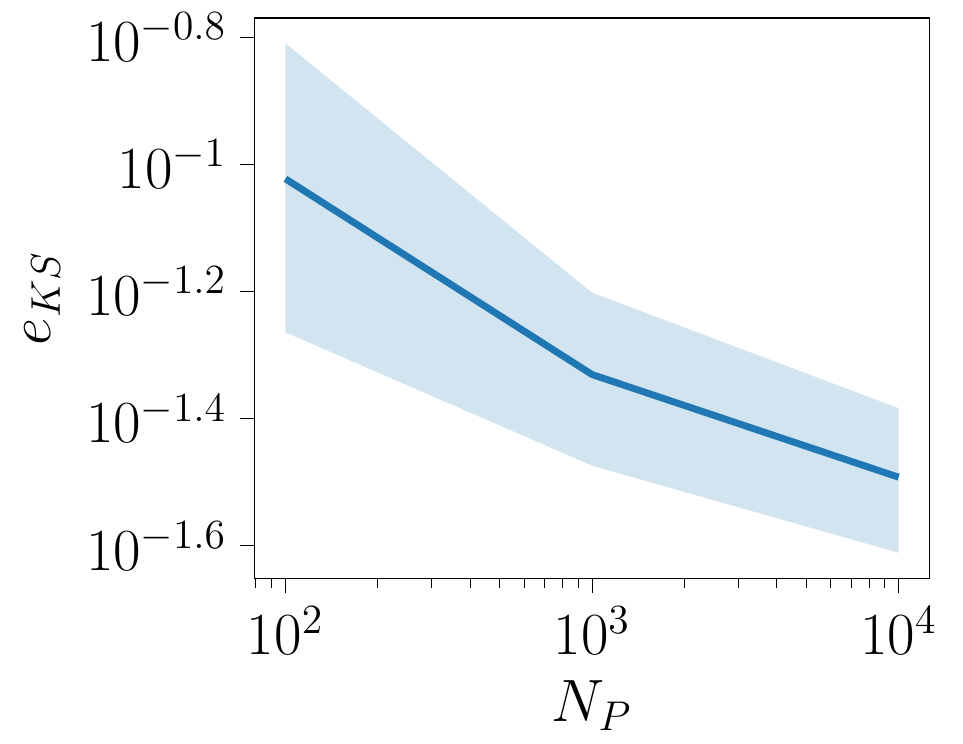}
	\end{subfigure}
	\begin{subfigure}{0.45\textwidth}\caption{}
	\centering
	\includegraphics[height=5cm]{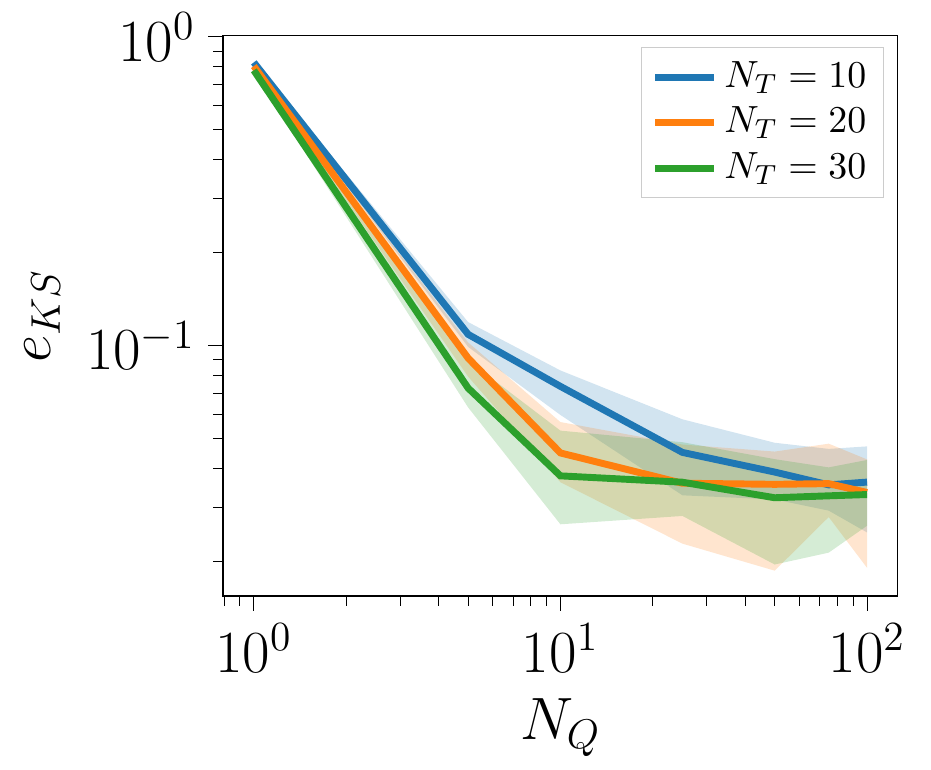}
	\end{subfigure}
    \caption {Three-bar truss with sliding-Gaussian material data, $M_e=10^5$. a) KS error {\sl vs.} population size, $N_T=20$ and $N_Q=100$. b) KS error {\sl vs.} number of quenches $N_Q$ for different number of trials $N_T$ using $N_P=5000$.
    The lines show the mean of $20$ repeated computations, the shaded area indicate the maximum deviation from mean.} \label{fig:Param_study}
\end{figure}

Fig.~\ref{fig:data_set_error}a shows the evolution of $\beta_e$ during the annealing process. As may be seen from the figure, $\beta_e$ increases monotonically with the number of quenches and tends towards a final value. This final value itself increases with the material data-set size as the average minimum distance between material data points decreases. Fig.~\ref{fig:data_set_error}b shows the dependence of the Kolmogorov–Smirnov error $e_{KS}$ on the number of material data points, which again exhibits a clear trend towards convergence. An analysis of the convergence plot reveals that, asymptotically for large material data sets, $e_{KS} \sim M_e^{-\alpha}$, where $\alpha \sim 0.36$ is the rate of convergence. These results are consistent with quantitative error estimates derived from analysis \cite{Conti:2022}. 
For completeness, 
Fig.~\ref{fig:Param_study}a shows the dependence of $e_{KS}$ on the population size $N_P$. Fig.~\ref{fig:Param_study}b shows the dependence on the number of quenches $N_Q$ for different number of trials $N_T$ using a population size of $N_P=5000$.

\subsection{Three-bar truss with Weibull tensile-strength distribution}\label{sec:three_bars_Weibull}

\begin{figure}[ht]
	\begin{subfigure}{0.45\textwidth}\caption{} \includegraphics[width=0.99\linewidth]{HW8_Problem1_Figure1.pdf}
	\end{subfigure}
    $\quad$
	\begin{subfigure}{0.40\textwidth}\caption{} \includegraphics[width=0.99\linewidth]{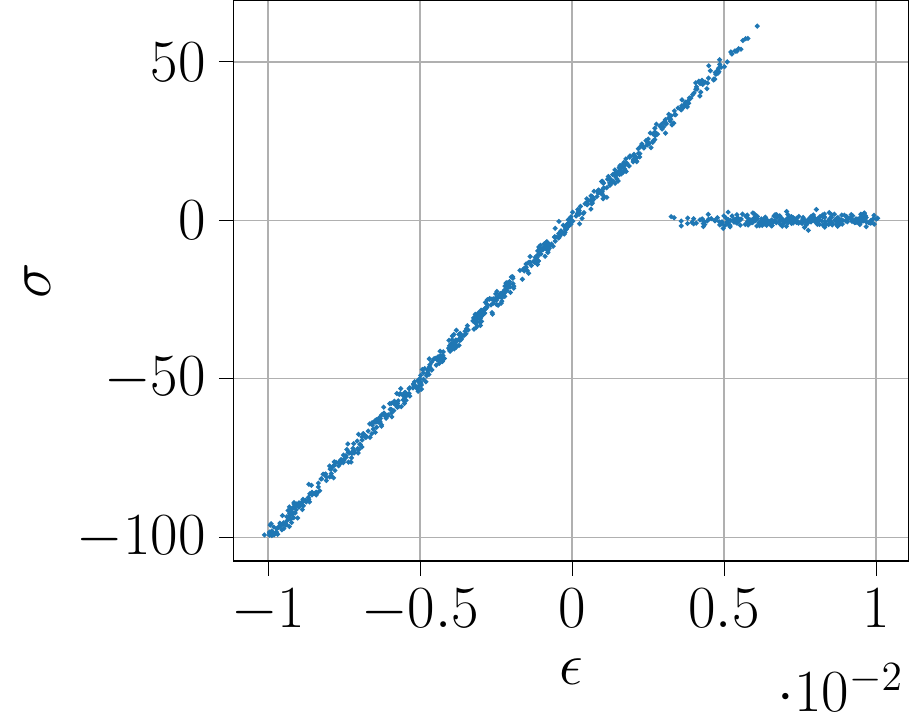}
	\end{subfigure}
    \caption{a) Truss with prescribed displacement $\Delta$ and reaction force $P$. b) Local material data set with $1000$ data points sampled from a bimodal likelihood function with failed and unfailed branches selected according to Weibull strength statistics, plus superimposed Gaussian noise.} \label{fig:dataset_fail}
\end{figure}

Next, we present a simple example that is designed to illustrate the ability of the method to deal effectively with complex data and behavior. The example concerns the same structure as in Section~\ref{sec:three_bars_Gaussian}, Fig.~\ref{fig:dataset_fail}a, but the material is now assumed to be brittle with random tensile strength.

The control parameter is now the displacement $\Delta$, which is increased monotonically. The outcome of interest is the corresponding the reaction force $P$. By virtue of the monotonicity of $\Delta$, the strains in all bars also increase monotonically. The behavior of the bars in compression is linear elastic. By contrast, in bars in tension the stress increases linearly up to the tensile strength $\sigma_t > 0$ of the material and subsequently drops to zero at larger strains. In addition, the tensile strength of the material is assumed to be random and to obey Weibull statistics, with distribution
\begin{equation}
    W(\sigma_t)
    =
    1-{\rm e}^{-(\sigma_t/\sigma_0)^p} ,
\end{equation}
where $\sigma_0$ and $p$ are material-specific parameters. Under these conditions, the likelihood of observing a local tensile state $(\epsilon_e,\sigma_e)$ is {\sl bimodal} and consists of: a branch $\sigma_e = \mathbb{C}_e \epsilon_e$  with likelihood $1-W(\mathbb{C}_e \epsilon_e)$, corresponding to an unfailed bar; and a branch $\sigma_e = 0$ with likelihood $W(\mathbb{C}_e \epsilon_e)$, corresponding to a failed bar.

A typical material data set sampled from this likelihood function, with superimposed Gaussian noise of standard deviation $s=10^{-4}$, is shown in Fig.~\ref{fig:dataset_fail}b. The data set is intended to mimic empirical data sets such as might be generated experimentally, including experimental scatter. Again, we emphasize that such material data sets are the only input to the calculations and, once generated, the underlying distribution whence they are sampled is jettisoned altogether and presumed unknown.

\begin{figure}[ht]
    \centering
	\begin{subfigure}{0.45\textwidth}\caption{}
    \centering
    \includegraphics[height=5.0cm]{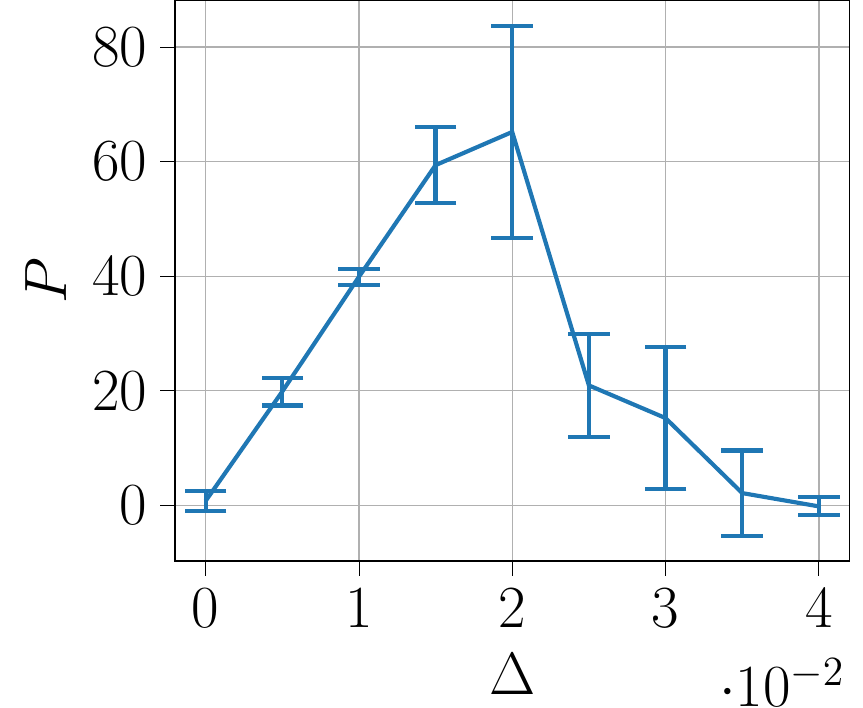}
	\end{subfigure}%
	\begin{subfigure}{0.45\textwidth}\caption{}
    \centering
    \includegraphics[height=5.0cm]{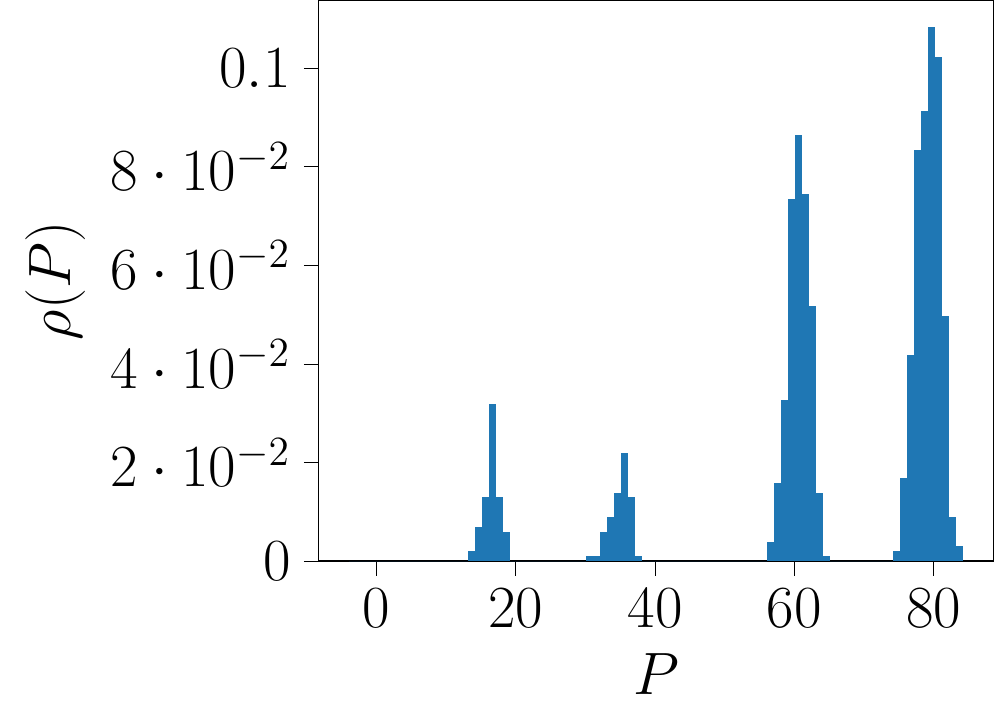}
	\end{subfigure}
    \caption{Three-bar truss with Weibull tensile strength data. a) Computed mean and standard deviation of the reaction force $P$ for prescribed displacement $\Delta$. b) Computed distribution of $P$ at $\Delta=2\times 10^{-2}$.} \label{fig:rf}
\end{figure}

For the simulation parameters, we choose $N_Q=50, N_T=10$ and $N_P=2\times 10^4$ with data-set size of $M_e=10^3$. Fig.~\ref{fig:rf}a shows the computed mean and standard deviation of the reaction force $P$ over a range of  prescribed displacements $\Delta$. Interestingly, the figure reveals three regimes: i) and initial linear elastic response with low uncertainty, as measured by the standard deviation; ii) an intermediate regime combining failed and unfailed bars characterized by a decreasing reaction and high uncertainty; and iii) full failure, or {\sl collapse}, with low uncertainty.

Fig.~\ref{fig:rf}b in turn shows the computed distribution of $P$ at $\Delta=2\times 10^{-2}$. Remarkably, the distribution is strongly {\sl multimodal}, a telltale hallmark of complexity. The various modes correspond to several different possible combinations of failed and unfailed bars and their attendant likelihoods. For the data used in calculations, the right bar is computed to remain unfailed with probability $1$ or, equivalently, the probability of the right bar failing is computed to be $0$. By contrast, the remaining two bars can be failed or unfailed with varying likelihoods: no failure (right peak), double failure (left peak), single failures (two middle peaks). From a global perspective, the various modes in the posterior likelihood function of $P$ correspond to a {\sl clustering}, or {\sl classification}, of global material data points close to the constraint set. The ability of the method to classify outcomes automatically and effectively bears remark.

\subsection{Lightweight space structure}\label{sec:pactruss_study}
\begin{figure}[ht]
    \includegraphics[width=0.40\linewidth]{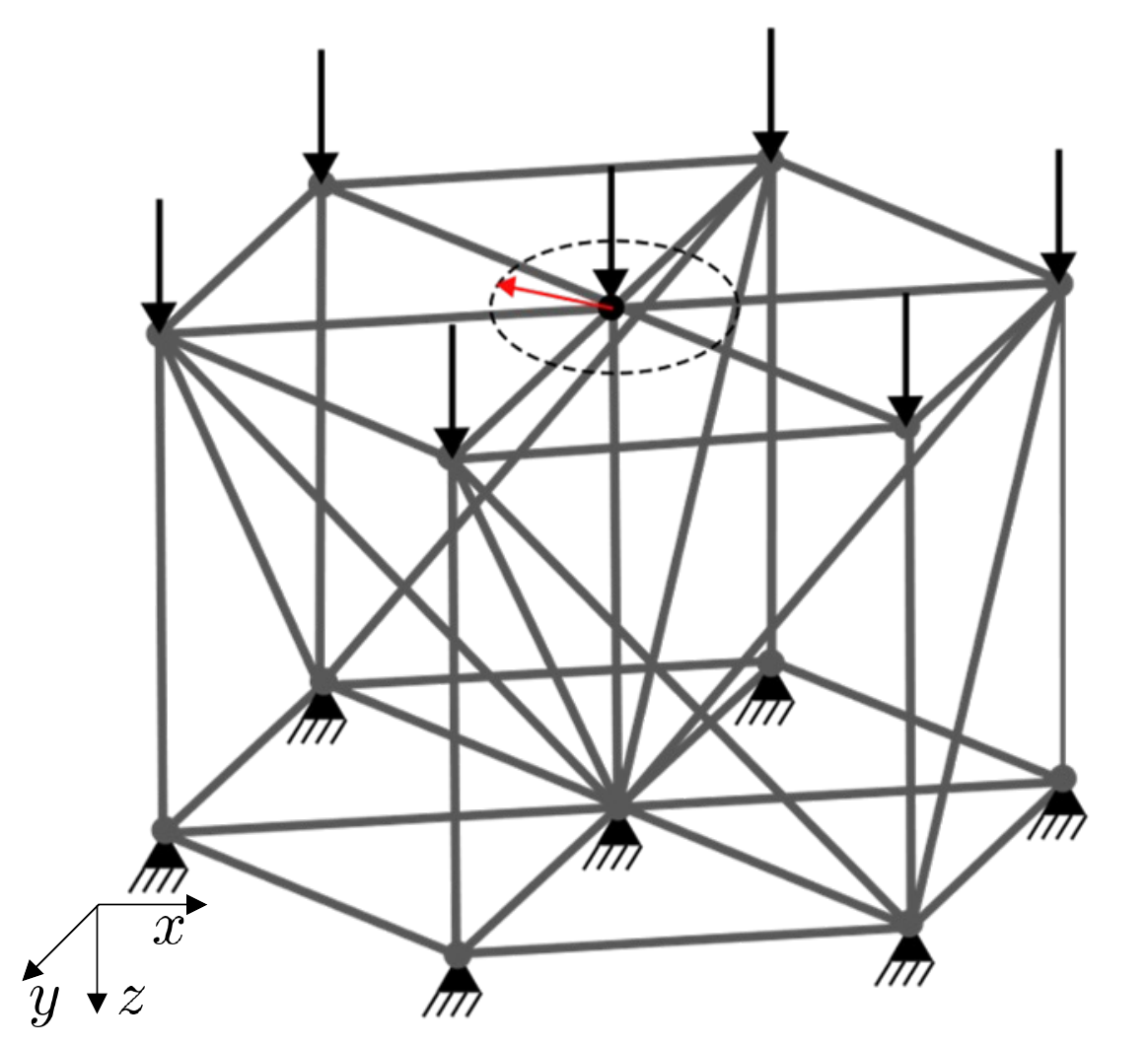}
    \caption{Lightweight space structure \cite{lee2016architecture}. The red arrow indicates the quantity of interest (joint eccentricity $u_e$) to be inferred from the analysis.} \label{fig:pactruss}
\end{figure}

By way of demonstration, we consider the lightweight structure shown in Fig.~\ref{fig:pactruss} designed to be part of a modular space telescope \cite{lee2016architecture}. We specifically aim to assess the feasibility of the proposed approach for higher-dimensional problems (constraint set dimension $N=39$) and its computational performance in terms of resource requirements and execution times.

In calculations, we set the cross-sectional area of the bars to $A=0.1$ and the length of the non-diagonal bars to $1$. The structure is constrained throughout its base and is loaded by longitudinal joint forces $P=5$ as shown in Fig.~\ref{fig:pactruss}. As a quantity of interest, we choose the top central joint eccentricity $u_e=\sqrt{u_x^2+u_y^2}$ whereby $u_x$ and $u_y$ are the in-plane displacements of the joint. We specifically consider a material data set with $M_e=10^4$ data points sampled from a sliding Gaussian as in Section~\ref{sec:three_bars_Gaussian}, cf.~Fig.~ \ref{fig:truss_gauss}. Conveniently, this assumption affords a closed-form analytical solution for the distribution of $u_e$ that can be used by way of reference, cf.~Section~\ref{suO4Ci}.

For algorithm parameters, we choose $N_P=2\times 10^3, N_T=10, N_Q=100$ and $ s_{0}=0.01$. For faster convergence to low-energy states, we initialize the population by means of $1000$ random min-dist solutions \cite{kirchdoerfer2016data} for noise-free data. As expected, the posterior distribution derived from the initial population is far from the exact distribution, Fig.~\ref{fig:pactruss_pdf_study}a. By contrast, a close agreement is achieved following the application of the population annealing algorithm, Fig.~\ref{fig:pactruss_pdf_study}b, which is remarkable in view of the relatively small population and material-data sizes used in the calculations.
{
For comparison, we note that the maximum-likelihood solution [15] is computed to be $u_{e,\rm ML}=0$. This comparison underscores that, while the maximum-likelihood solution can deal effectively with observational noise, it only identifies the most likely state of the system. By contrast, for intrinsically random systems it is often desirable to infer the entire posterior distribution of states, as in the present framework.

\begin{figure}[ht]
	\begin{subfigure}{0.5\textwidth}\caption{} \includegraphics[height=4.8cm]{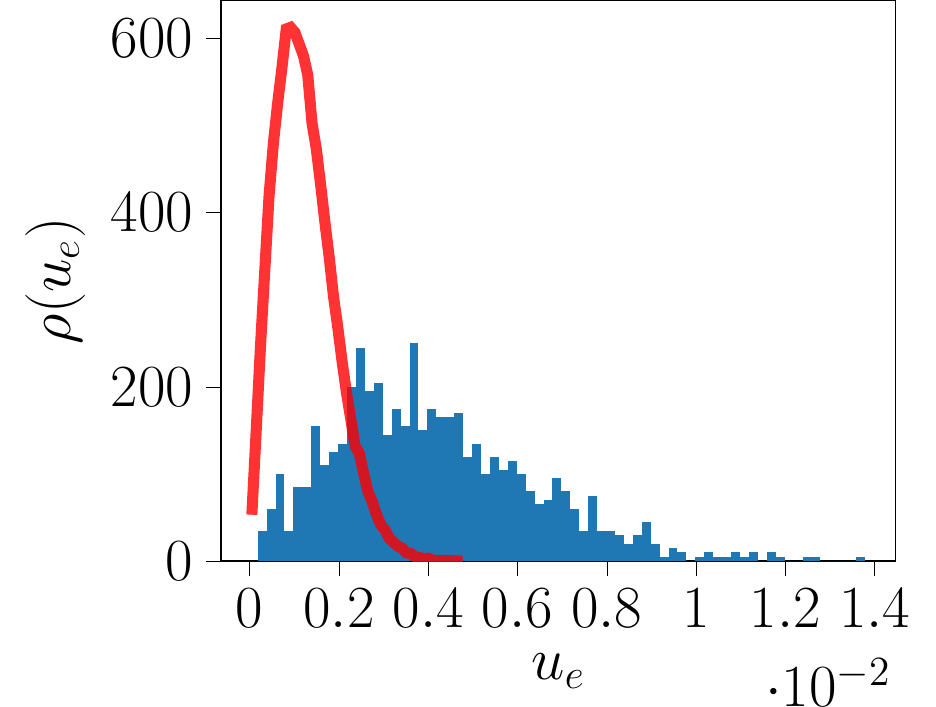}
	\end{subfigure}%
	\begin{subfigure}{0.5\textwidth}\caption{} \includegraphics[height=4.8cm]{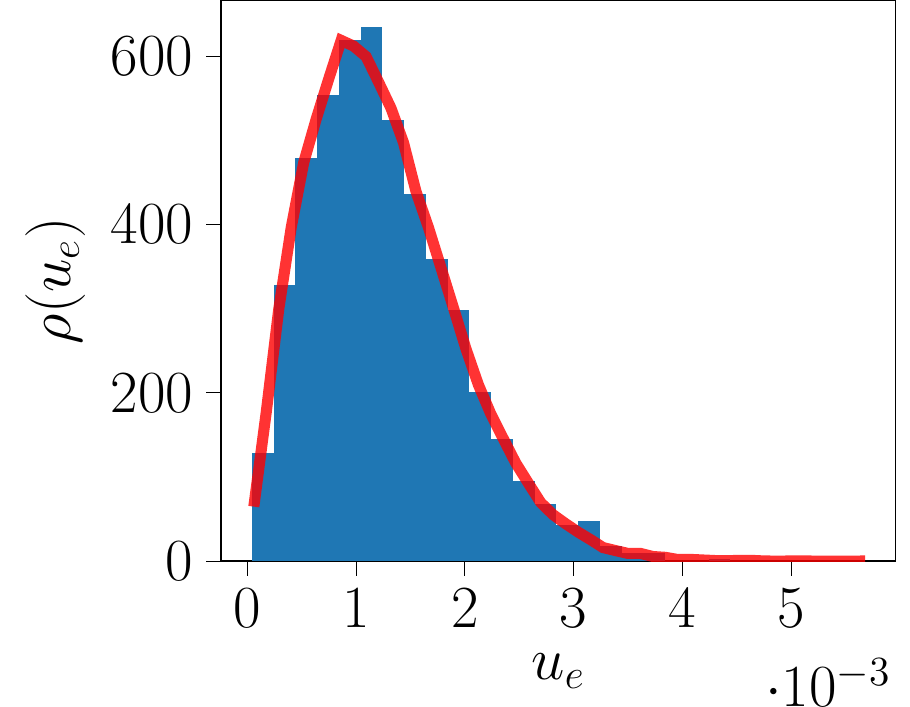}
	\end{subfigure}
    \caption {Lightweight space structure \cite{lee2016architecture}. Computed posterior distributions of the joint eccentricity $u_e$ and exact reference distribution (red curves). a) Posterior distribution obtained from an initial population of $1000$ random min-dist solutions. b) Posterior distribution following application of the population annealing algorithm. }
    \label{fig:pactruss_pdf_study}
\end{figure}

Finally, we assess the impact of the approximate nearest neighbor search on accuracy and CPU times for the population size $N_P=2\times 10^3$. The absolute CPU times are, of course, implementation and machine dependent. In this work, the computations were performed on a 12-Core AMD Ryzen 9 3900X machine with Python bindings \cite{jakob2019pybind11} to a C++ implementation using Standard Template Library Parallel Algorithms.

The CPU times for a direct evaluation of Eq. (\ref{eq:energy}) without any data structuring or acceleration is $\sim 662$ seconds. The radius search with full backtracking terminates in $\sim 92$ seconds, or a seven-fold acceleration. Fig.~\ref{fig:pactruss_cpu_study} shows the effect of the degree of backtracking measured in terms of $N_{\rm checks}$. We see from the figure that, whereas the computation time increases rapidly with the degree of backtracking, accuracy is not substantially improved above $N_{\rm checks}= 2^3$, at which limit the CPU time is reduced to $~8$ seconds. This tradeoff further suggests that the performance of the algorithm can benefit greatly from data structuring techniques such as approximate nearest-neighbor searches \cite{eggersmann2021efficient}, especially for large systems and material data sets. Furthermore, we recall that the algorithm is embarrassingly parallel, which opens the way for efficient parallel implementations, e.~g., on GPU machines \cite{barash2017gpu}.

\begin{figure}[ht]
    \centering
	\begin{subfigure}{0.45\textwidth}\caption{}
    \centering
    \includegraphics[height=4.5cm]{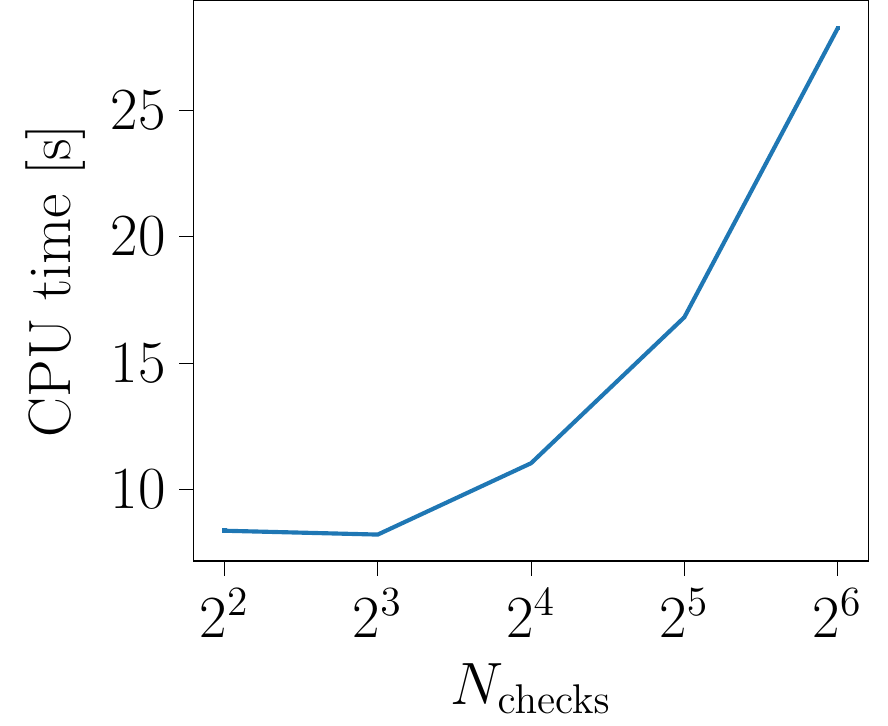}
	\end{subfigure}%
	\begin{subfigure}{0.45\textwidth}\caption{}
    \centering
    \includegraphics[height=4.5cm]{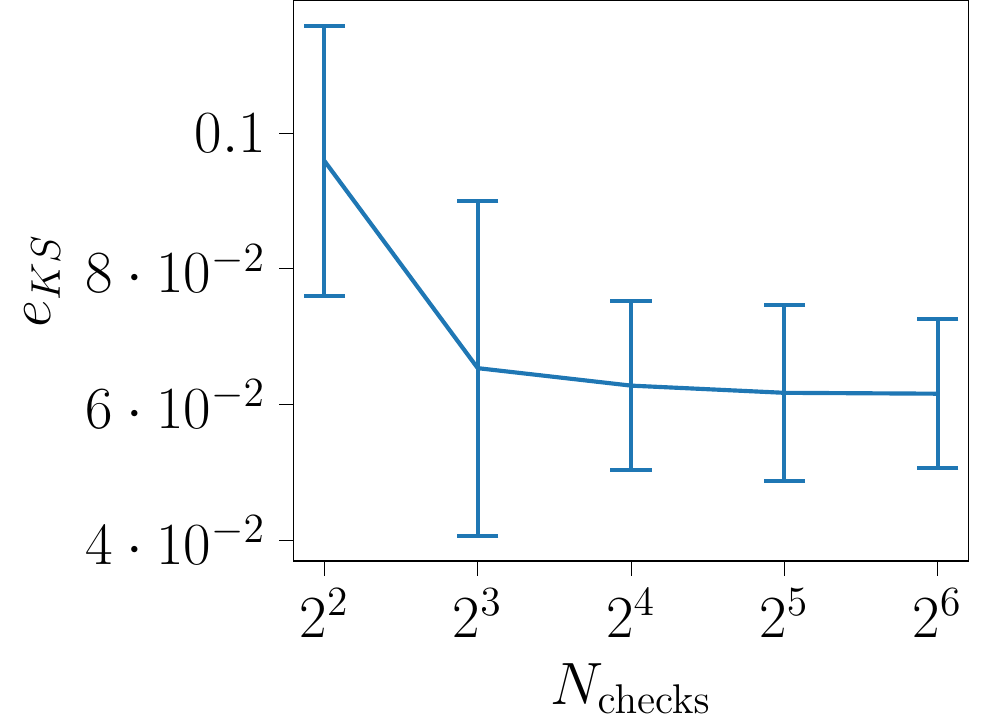}
	\end{subfigure}
    \caption {Lightweight space structure \cite{lee2016architecture}. Influence of the number of checks for backtracking approximate in the nearest-neighbor search. a) CPU times in seconds on a 12-Core AMD Ryzen 9 3900X machine. b) Error $e_{KS}$ of node eccentricity $u_e$ with error bars derived from a sample of $25$ calculations.}
    \label{fig:pactruss_cpu_study}
\end{figure}

\section{Summary and concluding remarks}

The behavior of many materials is characterized by a {\sl likelihood measure} instead of a constitutive relation. Specifically, the material likelihood measure expresses the likelihood of observing a material state, understood as a point in stress-strain space, or {\sl phase space}, in the laboratory. We have assumed that the material likelihood measure is known only partially through an empirical point-data set in phase space. The state of the solid or structure is additionally subject to compatibility and equilibrium constraints that encode, in particular, all information pertaining to geometry, boundary conditions and loading. The problem is then to infer the {\sl posterior likelihood} of a given outcome of interest.

We have presented an {\sl ansatz}-free Data-Driven method of inference that determines likelihoods of outcomes directly from the empirical material data. In particular, the method of inference requires no material or prior modeling. The computation of expectations is reduced to explicit sums over points in local material data sets and quadratures over admissible states, i.~e., states satisfying compatibility and equilibrium. The method reduces to the max-ent Data-Driven approach proposed in
\cite{kirchdoerfer2017data} when the analysis is restricted to the computation of maximum-likelihood outcomes. In its present form, the Data-Driven method of inference delivers a full characterization of expectations and distributions of outcomes from general material data sets.

The complexity of the Data-Driven inference calculations is linear in the number of material data points and in the number of members in the structure or Gauss points in the solid. Efficient population annealing procedures and fast-search algorithms for accelerating calculations have been presented. Specifically, the population annealing algorithm reduces the complexity of the calculations and it renders it linear in the size of the material data-point population. The calculations are further accelerated by recourse to a hierarchical $k$-means algorithm for the computation of the annealing energy with limits on the number backtracking operations.

The convergence properties of the method with respect to the empirical data have been assessed with the aid of a standard verification test based on Gaussian material data. Robust convergence of posterior distributions is obtained, consistent with quantitative error estimates derived from analysis \cite{Conti:2021, Conti:2022}. It bears emphasis that, in these tests and in all subsequent calculations, the empirical material-data sets are the only input to the calculations and, once generated, the underlying distribution whence they are sampled is discarded altogether and plays no subsequent role.

The full power of the method becomes apparent in applications to strongly non-Gaussian material data sets. Thus, we have presented an example concerned with brittle materials exhibiting random tensile strength. Specifically, the tensile strength of the material is assumed to obey Weibull statistics. The computed posterior distributions in this case are strongly multimodal and encode complex behavior characterized by multiple alternatives. It bears emphasis that the material data-sets arising in this example challenge conventional methods in that: i) there is no reduced manifold in stress-strain space, and ii) the prior distribution is strongly non-Gaussian. Indeed, owing to the random character of the tensile strength of the solid, the feasible material region in stress-strain space is not a manifold, but a full set. In addition, in Bayesian approaches to inference it is common to hypothesize Gaussian priors for the observational noise as a matter of convenience. Evidently, such  hypothesis is not born out in the Weibull strength statistics scenario of the current example and may lead to lack or convergence or, worse still, to convergence to the wrong limit (cf.~\cite{Owhadi:2015}). Against this backdrop, the ability of the proposed Data-Driven method of inference to deal effectively with general material data sets and complex behavior without need for models, hypothesis or assumptions bears remark.

Finally, we have presented a simple benchmark case, concerned with the deformations of a lightweight space structure, that demonstrates the expected computational performance of the method. The calculations attest to the effectiveness of acceleration methods such as population annealing and hierarchical $k$-means searches. While the benchmark calculations presented here are of modest size, the linear scaling of the algorithm with system size and population size, as well as the embarrassingly parallel property of the algorithm, bode well for applications at scale.

\section{Acknowledgements}
All authors gratefully acknowledge the financial support of the Deutsche Forschungsgemeinschaft (DFG) and French Agence Nationale de la Recherche (ANR) through the project "Direct Data-Driven Computational Mechanics for Anelastic Material Behaviours" (ANR-19-CE46-0012-01, RE 1057/47-1, project number 431386925) within the French-German Collaboration for Joint Projects in Natural, Life and Engineering (NLE) Sciences. SR gratefully acknowledges the funding of the DFG projects "Model order reduction in space and parameter dimension - towards damage-based modeling of polymorphic uncertainty in the context of robustness and reliability" (SPP 1886, project number 312911604) and "Coupling of intrusive and non-intrusive locally decomposed model order reduction techniques for rapid simulations of road systems" (TP B05/TRR 339, project number 453596084). MO is grateful for support from the Deutsche Forschungsgemeinschaft (DFG, German Research Foundation) {\sl via} project 211504053 - SFB 1060; project 441211072 - SPP 2256; and project 390685813 -  GZ 2047/1 - HCM.
\bibliographystyle{unsrt}
\bibliography{biblio}

\begin{thebibliography}{10}

\bibitem{Bower:2010}
A.~F. Bower.
\newblock {\em Applied Mechanics of Solids}.
\newblock CRC Press, Boca Raton, Fla., 2010.

\bibitem{Reynolds:1974}
W.~N. Reynolds and J.~V. Sharp.
\newblock Crystal shear limit to carbon fibre strength.
\newblock {\em Carbon}, 12:103--–110, 1974.

\bibitem{Bennett:1983}
S.~C. Bennett, D.~J. Johnson, and W.~Johnson.
\newblock Strength-structure relationships in pan-based carbon fibres.
\newblock {\em Journal of Material Science}, 18:3337--3347, 1983.

\bibitem{Owhadi:2015}
H.~Owhadi, C.~Scovel, and T.~Sullivan.
\newblock On the brittleness of bayesian inference.
\newblock {\em SIAM REVIEW}, 57(4):566--582, 2015.

\bibitem{Dashti:2017}
Masoumeh Dashti and Andrew~M. Stuart.
\newblock {\em The Bayesian Approach to Inverse Problems}, pages 311--428.
\newblock Springer International Publishing, Cham, 2017.

\bibitem{Knapik:2011}
B.~T. Knapik, A.~W. van~der Vaart, and J.~H. van Zanten.
\newblock Bayesian inverse problems with gaussian priors.
\newblock {\em Annals of Statistics}, 39:2626--2657, 2011.

\bibitem{Knapik:2016}
B.~T. Knapik, B.~T. Szab\'o, A.~W. van~der Vaart, and J.~H. van Zanten.
\newblock Bayes procedures for adaptive inference in inverse problems for the
  white noise model.
\newblock {\em Probab. Theory Relat. Fields}, 164(3--4):771--813, 2016.

\bibitem{Bader:1993}
M.~G. Bader, K.~L. Pickering, A.~Buxton, A.~Rezaifard, and P.~A. Smith.
\newblock Failure micromechanisms in continuous carbon-fibre/epoxy-resin
  composites.
\newblock {\em Composites Science and Technology}, 48:135--142, 1993.

\bibitem{Naito:2008}
K.~Naito, Y.~Tanaka, J.-M. Yang, and Y.~Kagawa.
\newblock Tensile properties of ultrahigh strength pan-based, ultrahigh modulus
  pitch-based and high ductility pitch-based carbon fibers.
\newblock {\em Carbon}, 46:189--195, 2008.

\bibitem{Conti:2021}
S.~Conti, F.~Hoffmann, and M.~Ortiz.
\newblock Model-free data-driven inference.
\newblock {\em arXiv.org}, math:arXiv:2106.02728, 2021.

\bibitem{Kullback:1951}
S.~Kullback and R.~A. Leibler.
\newblock On information and sufficiency.
\newblock {\em Annals of Mathematical Statistics}, 22(1):79--86, 1951.

\bibitem{Kullback:1959}
S~Kullback.
\newblock {\em Information Theory and Statistics}.
\newblock John Wiley \& Sons, New-York, 1959.

\bibitem{Pinski:2015}
F.~J. Pinski, G.~Simpson, A.~M. Stuart, and H.~Weber.
\newblock Kullback-leibler approximation for probability measures on infinite
  dimensional spaces.
\newblock {\em SIAM Journal on Mathematical Analysis}, 47(6):4091--4122, 2015.

\bibitem{kirchdoerfer2016data}
Trenton Kirchdoerfer and Michael Ortiz.
\newblock Data-driven computational mechanics.
\newblock {\em Computer Methods in Applied Mechanics and Engineering},
  304:81--101, 2016.

\bibitem{kirchdoerfer2017data}
Trenton Kirchdoerfer and Michael Ortiz.
\newblock Data driven computing with noisy material data sets.
\newblock {\em Computer Methods in Applied Mechanics and Engineering},
  326:622--641, 2017.

\bibitem{Conti:2022}
Sergio Conti, Franca Hoffmann, and Michael Ortiz.
\newblock Convergence rates for ansatz-free data-driven inference in physically
  constrained problems.
\newblock {\em arXiv preprint arXiv:2210.02846}, 2022.

\bibitem{kanno2019mixed}
Yoshihiro Kanno.
\newblock Mixed-integer programming formulation of a data-driven solver in
  computational elasticity.
\newblock {\em Optimization Letters}, 13(7):1505--1514, 2019.

\bibitem{metropolis1953equation}
Nicholas Metropolis, Arianna~W Rosenbluth, Marshall~N Rosenbluth, Augusta~H
  Teller, and Edward Teller.
\newblock Equation of state calculations by fast computing machines.
\newblock {\em The journal of chemical physics}, 21(6):1087--1092, 1953.

\bibitem{iba2001population}
Yukito Iba.
\newblock Population monte carlo algorithms.
\newblock {\em Transactions of the Japanese Society for Artificial
  Intelligence}, 16(2):279--286, 2001.

\bibitem{machta2010population}
Jonathan Machta.
\newblock Population annealing with weighted averages: A monte carlo method for
  rough free-energy landscapes.
\newblock {\em Physical Review E}, 82(2):026704, 2010.

\bibitem{weigel2021understanding}
Martin Weigel, Lev Barash, Lev Shchur, and Wolfhard Janke.
\newblock Understanding population annealing monte carlo simulations.
\newblock {\em Physical Review E}, 103(5):053301, 2021.

\bibitem{wang2015comparing}
Wenlong Wang, Jonathan Machta, and Helmut~G Katzgraber.
\newblock Comparing monte carlo methods for finding ground states of ising spin
  glasses: Population annealing, simulated annealing, and parallel tempering.
\newblock {\em Physical Review E}, 92(1):013303, 2015.

\bibitem{barash2017gpu}
Lev~Yu Barash, Martin Weigel, Michal Borovsk{\`y}, Wolfhard Janke, and Lev~N
  Shchur.
\newblock Gpu accelerated population annealing algorithm.
\newblock {\em Computer Physics Communications}, 220:341--350, 2017.

\bibitem{eggersmann2021efficient}
Robert Eggersmann, Laurent Stainier, Michael Ortiz, and Stefanie Reese.
\newblock Efficient data structures for model-free data-driven computational
  mechanics.
\newblock {\em Computer Methods in Applied Mechanics and Engineering},
  382:113855, 2021.

\bibitem{fukunaga1975branch}
Keinosuke Fukunaga and Patrenahalli~M. Narendra.
\newblock A branch and bound algorithm for computing k-nearest neighbors.
\newblock {\em IEEE transactions on computers}, 100(7):750--753, 1975.

\bibitem{muja2014scalable}
Marius Muja and David~G Lowe.
\newblock Scalable nearest neighbor algorithms for high dimensional data.
\newblock {\em IEEE transactions on pattern analysis and machine intelligence},
  36(11):2227--2240, 2014.

\bibitem{graves2011automatic}
Todd~L Graves.
\newblock Automatic step size selection in random walk metropolis algorithms.
\newblock {\em arXiv preprint arXiv:1103.5986}, 2011.

\bibitem{lee2016architecture}
Nicolas~N Lee, Joel~W Burdick, Paul Backes, Sergio Pellegrino, Kristina
  Hogstrom, Christine Fuller, Brett Kennedy, Junggon Kim, Rudranarayan
  Mukherjee, Carl Seubert, et~al.
\newblock Architecture for in-space robotic assembly of a modular space
  telescope.
\newblock {\em Journal of Astronomical Telescopes, Instruments, and Systems},
  2(4):041207, 2016.

\bibitem{jakob2019pybind11}
Wenzel Jakob, Jason Rhinelander, and Dean Moldovan.
\newblock pybind11--seamless operability between c++ 11 and python, 2017.
\newblock {\em URL https://github. com/pybind/pybind11}, 2019.

\end{thebibliography}

\end{document}